\documentclass{aa}

\usepackage{graphicx}
\usepackage{txfonts}
\usepackage{textcomp}
\usepackage{natbib}
\bibpunct{(}{)}{;}{a}{}{,} 

\begin{document}
    \title{On the sensitivity of gravito-inertial modes to differential rotation in intermediate-mass main-sequence stars}
   \titlerunning{Differential rotation}

   \author{T.~Van Reeth \inst{1,2,3},
   J.~S.~G.~Mombarg \inst{4,1},
   S.~Mathis\inst{5},
   A.~Tkachenko\inst{1},
   J.~Fuller\inst{6,2},
   D.~M.~Bowman\inst{1},
   B.~Buysschaert\inst{1,7},
   C.~Johnston\inst{1},
   A.~Garc\'ia Hern\'andez\inst{8,9},
   J.~Goldstein\inst{10},
   R.~H.~D.~Townsend\inst{10},
   C.~Aerts\inst{1,4}}
   
   \institute{Institute of Astronomy, KU Leuven, Celestijnenlaan 200D, B-3001 Leuven, Belgium
   \and Kavli Institute for Theoretical Physics, University of California, Santa Barbara, CA 93106, USA
   \and Sydney Institute for Astronomy (SIfA), School of Physics, University of Sydney, New South Wales 2006, Australia
   \and Department of Astrophysics, IMAPP, Radboud University Nijmegen, PO Box 9010, 6500 GL Nijmegen, The Netherlands
   \and Laboratoire AIM Paris-Saclay, CEA/DRF-Universit\'e Paris Diderot-CNRS, IRFU/SAp Centre de Saclay, 91191 Gif-sur-Yvette, France
   \and TAPIR, Walter Burke Institute for Theoretical Physics, Mailcode 350-17, Caltech, Pasadena, CA 91125, USA
   \and LESIA, Observatoire de Paris, PSL Research University, CNRS, Sorbonne Universit\'es, UPMC Univ. Paris 6, Univ. Paris Diderot,
Sorbonne Paris Cit\'e, France
   \and Department of Theoretical Physics and Cosmology, University of Granada (UGR), E-18071 Granada, Spain
   \and Instituto de Astrof\'isica e Ci\^encias do Espa\c{c}o, Universidade do Porto, CAUP, Rua das Estrelas, P-4150-762 Porto, Portugal
   \and Department of Astronomy, University of Wisconsin-Madison, 475 Charter Street, Madison, WI 53706, USA}
   
   \authorrunning{T. Van Reeth et al.}
   
   \date{Received; accepted}
 
  \abstract
   {While rotation has a major impact on stellar structure and evolution, its effects are not well understood. Thanks to high-quality and long timebase photometric observations obtained with recent space missions, we are now able to study stellar rotation more precisely.}
   {We aim to constrain radial differential rotation profiles in $\gamma$\,Doradus ($\gamma$\,Dor) stars, and to develop new theoretical seismic diagnosis for such stars with rapid and potentially non-uniform rotation.}
   {We derive a new asymptotic description which accounts for the impact of weak differential near-core rotation on gravity-mode period spacings. The theoretical predictions are illustrated from pulsation computations with the code GYRE and compared with observations of $\gamma$\,Dor stars. When possible, we also derive the surface rotation rates in these stars by detecting and analysing signatures of rotational modulation, and compute the core-to-surface rotation ratios.}
   {Stellar rotation has to be strongly differential before its effects on period spacing patterns can be detected, unless multiple period spacing patterns can be compared. Six stars in our sample exhibit a single unexplained period spacing pattern of retrograde modes. We hypothesise that these are Yanai modes. Finally, we find signatures of rotational spot modulation in the photometric data of eight targets.}
   {If only one period spacing pattern is detected and analysed for a star, it is difficult to detect differential rotation. A rigidly rotating model will often provide the best solution. Differential rotation can only be detected when multiple period spacing patterns have been found for a single star or its surface rotation rate is known as well. This is the case for eight stars in our sample, revealing surface-to-core rotation ratios between 0.95 and 1.05.}


   \maketitle

\section{Introduction}
\label{sec:intro}

Rotation is a crucial element of stellar structure and evolution \citep[e.g.,][and references therein]{Maeder2009}. It induces and influences various types of angular momentum transport and chemical mixing processes inside the star, such as shear mixing \citep[e.g.,][]{Heger2000,Mathis2004,Maeder2013}, meridional circulation \citep[e.g.,][]{Zahn1992,Maeder1998,Mathis2004b} and waves \citep[e.g.,][]{Lee1993,TalonCharbonnel1998,TalonCharbonnel2005,Mathis2009,Rogers2015}. These lead to additional transport of fuel into the nuclear burning regions, influencing the nucleosynthesis and extending the life of the star \citep[e.g.][]{Zahn1992,Pinsonneault1997,Maeder2000b}. The mixing processes also have a potential impact on the element abundances at the stellar surface \citep[e.g.,][]{Brott2011a,Brott2011b}. Furthermore, stellar rotation has a strong influence on stellar magnetism \citep[e.g.,][]{Mestel1999,Spruit1999} and vice versa \citep[e.g.,][]{Mestel1988,Spruit1999,Mathis2005}. The centrifugal acceleration distorts the shape of the star \citep[e.g.,][]{Roxburgh2004,Jackson2005,Roxburgh2006,MacGregor2007,Espinosa2013}, and the distribution of angular momentum influences the convective stability inside the star \citep[e.g.,][]{Maeder2013}. 

The full impact of stellar rotation is far from understood \citep[e.g.,][]{Aerts2014,Cazorla2017b}. While there have been considerable efforts to do multi-dimensional stellar modelling \citep[e.g.,][]{Rieutord2016}, these often have a high computational cost. Hence, state-of-the-art stellar evolution codes typically use one-dimensional stellar models  in which higher dimensional processes such as stellar rotation are averaged over isobars \citep[e.g.,][]{Chaboyer1995, Heger2000, Eggenberger2008, Ekstrom2012, Chieffi2013, Paxton2013, Marques2013, Palacios2003, Charbonnel2010, Amard2016}. This approach is often incomplete, and potentially numerically unstable \citep[e.g.,][]{Lau2014,Lattanzio2015,Rieutord2016,Edelmann2017}. 


Thanks to recent photometric space missions such as MOST \citep{Walker2003}, CoRoT \citep{Auvergne2009}, {\em Kepler} \citep{Borucki2010,Koch2010}, K2 \citep{Howell2014}, and BRITE \citep{Weiss2014,Pablo2016}, there has been significant progress in constraining stellar rotation using observations. These high-quality high-cadence and long timebase data have led to major advances in asteroseismology, i.e., the study of the internal stellar structure, dynamics and evolution by analysing non-radial stellar pulsations. This has been particularly successful for low-mass main-sequence stars with radiative cores and convective envelopes \citep[e.g.,][]{Benomar2015,vanSaders2016}, subgiants \citep[e.g.,][]{Deheuvels2014}, and red giants \citep[e.g.,][]{Beck2012,Deheuvels2012,Ceillier2017}. However, asteroseismology is also a valuable tool to study intermediate- to high-mass main sequence stars with convective cores and radiative envelopes. 

Gamma Doradus ($\gamma$\,Dor) stars, with $1.4\,M_\odot \lesssim M_* \lesssim 2.0\,M_\odot$, and slowly-pulsating B-type (SPB) stars, with $2.5\,M_\odot \lesssim M_* \lesssim 8\,M_\odot$, exhibit high-order gravity-mode (g-mode) pulsations, gravito-inertial pulsations \citep{VanReeth2016}, and/or purely inertial pulsations, such as r-modes \citep{Saio2018}. The restoring forces for gravity-modes and purely inertial pulsation modes are buoyancy and the Coriolis force, respectively. In the case of gravito-inertial pulsation modes, both forces contribute. As predicted by asymptotic theory, the pulsation periods for $\gamma$\,Dor and SPB stars were observed to form period spacing patterns \citep[e.g.,][]{Chapellier2012,Chapellier2013,Kurtz2014,Bedding2015,Saio2015,VanReeth2015,Ouazzani2017}. The pulsation periods are equidistant in the asymptotic regime (with radial order $n$ $\gg$ spherical degree $l$) for a non-rotating chemically homogeneous star \citep{Tassoul1980}. Chemical gradients in the deep stellar interior cause pulsation mode trapping, which introduces non-uniform variations in the spacings \citep{Miglio2008a}. On the other hand, the stellar rotation leads to shifts in the observed pulsation mode frequencies \citep{Bouabid2013,Salmon2014,VanReeth2015, Moravveji2016}. For slowly rotating stars the observed pulsation modes are split into frequency multiplets that depend on the mode identification. For moderate to fast rotators, i.e., with rotation on the order of or more than 20\% of the critical rotation rate, the observed period spacing patterns have a clear slope \citep[e.g.,][]{VanReeth2016, Ouazzani2017}. Prograde (azimuthal order $m > 0$) and zonal modes have a downward slope, i.e., the spacing between consecutive pulsation periods decreases with increasing pulsation period, i.e., radial order $n$. The period spacing patterns of retrograde modes (with $m < 0$) mostly have an upward slope. For stars with detected period spacing patterns, this has been exploited to derive the near-core stellar rotation \citep[e.g.,][]{Kurtz2014,Saio2015,Triana2015,Murphy2016,SchmidAerts2016,VanReeth2016,Ouazzani2017,Saio2018}. 

It is often assumed that these stars are (quasi-)rigidly rotating. Notable exceptions are {\em (i)} the analysis of the slowly rotating B-type star KIC\,10526294 \citep{Triana2015}, whereby the impact of differential rotation on the pulsation mode splitting was studied, or {\em (ii)} the analysis of hybrid $\gamma$\,Dor/$\delta$\,Sct pulsators that exhibit both g-mode and p-mode pulsations \citep[e.g.,][]{Kurtz2014}. If the p-mode pulsation spectrum is sufficiently regular, it can be analysed to derive the surface rotation rate \citep[e.g.,][]{Kurtz2014,Saio2015,SchmidAerts2016,Paparo2016a,Paparo2016b}. 

In this work, we aim to constrain possible radial differential rotation in $\gamma$\,Dor stars. We study the problem using two different approaches. First, we take a closer look at the theoretically expected effects of differential rotation in the near-core regions on the g-mode pulsations (Section \ref{subsec:theor} and Appendix \ref{Appendix:theor}), and use them to place a constraint on the differential rotation of observed $\gamma$\,Dor stars (Section \ref{subsec:sample}). Second, we determine the core-to-surface rotation ratio for stars in the same sample using rotational spot modulation (Section \ref{sec:rotmod}) to constrain the stellar surface rotation rate. Finally, we will discuss these results, and draw our conclusions (Section \ref{sec:conclusions}).

\section{Differential rotation in the near-core regions}
\label{sec:diffrot_core}

\subsection{Theoretical description of the g-modes}
\label{subsec:theor}
Computing the complete impact of stellar rotation on pulsations is computationally expensive \citep{Dintrans2000,Ballot2010}. A common method to reduce this computational cost for g-mode pulsations is to use the traditional approximation of rotation (TAR), whereby {\em (i)} the horizontal component of the stellar rotation vector in the equation of motion is ignored, and {\em (ii)} the studied star is assumed to be spherically symmetric \citep[e.g., ][and Appendix \ref{Appendix:theor}]{Eckart1960,Friedlander1987,Lee1987,Lee1997,Townsend2005,Mathis2009}. For slow to moderate rotators where the stratification restoring force dominates the Coriolis acceleration along the radial direction, these assumptions are reasonable. The horizontal component of the pulsation displacement vector (and velocity) is dominant for g-modes, which are most sensitive to the near-core regions in the star where the deformation of the stellar structure caused by the centrifugal force is minimal.

When the star is assumed to be rigidly rotating, the g-mode pulsation periods in the corotating reference frame are given by
\begin{equation}
 P_{\rm co} = \frac{\Pi_0}{\sqrt{\lambda_{\nu km}}}\left(n + \alpha_g\right), \label{eq:Pg_uni}
\end{equation}
with
\begin{equation}
 \Pi_0 = \frac{2\pi^2}{\int_{r_1}^{r_2}\frac{N}{r}\mathrm{d}r}, \label{eq:pi0}
\end{equation}
as described by \citet{Bouabid2013}. Here, $N$ is the stable stratification Brunt-V\"ais\"al\"a frequency defined in Appendix \ref{Appendix:theor}, $\lambda_{\nu km}$ is the eigenvalue of the Laplace Tidal Equation, and $\nu$ the corresponding spin parameter given by $\nu = 2f_{\rm rot}P_{\rm co}$, with $f_{\rm rot}=\displaystyle{\frac{\Omega}{2\pi}}$ the rotation frequency ($\Omega$ being the rotation angular velocity). The integers $k$ and $m$ provide the pulsation mode identification as defined by \citet{Lee1997}, i.e., $k = l - |m|$ for g-modes, whereby $l$ is the spherical degree of the pulsation mode, and $k < 0$ for purely inertial modes; $n$ is the integer corresponding to the quantification along the radial direction. $\alpha_g$ is a phase term dependent on the stellar structure at the boundaries of the pulsation mode cavity. For stars with a convective core and  convective envelope, $\alpha_g \simeq 0.5$. The turning points of the pulsation mode cavity are indicated by $r_1$ and $r_2$.

Eq. (\ref{eq:Pg_uni}) can be generalised for purely radial differential rotation (the so-called shellular rotation), where the angular velocity is a function of $r$ only. Using the asymptotic theory for the propagation of low-frequency gravito-inertial modes in differentially rotating stars developed by \cite{Mathis2009} \citep[see also][]{Ogilvie2004} and detailed in Appendix \ref{Appendix:theor}, one can show that
\begin{equation}2\pi^2\left(n+\alpha_g\right) = \int_{r_1}^{r_2}\frac{\sqrt{\lambda_{\nu km}\left(r\right)}}{f_{\rm in}-mf_{\rm rot}\left(r\right)}\frac{N\left(r\right)}{r}\mathrm{d}r,\label{eq:Pg_diff}\end{equation}
where $f_{\rm in}$ is the frequency of the pulsation with mode identification $(n,k,m)$ as seen by the observer in the inertial reference frame. Here, both the eigenvalue $\lambda_{\nu km}\left(r\right)$ and the rotation profile $f_{\rm rot}\left(r\right)$ vary as a function of the radial coordinate $r$, and are included within the integral. 

We see from Eq. (\ref{eq:Pg_diff}) that small variations in the stellar rotation profile or in the Brunt-V\"ais\"al\"a frequency profile have a comparable effect on the pulsation frequencies. This can be illustrated by the case of a weak radial differential rotation defined as in \cite{Mathis2008}
\begin{equation}
f_{\rm rot}(r)=f_{\rm uni}+\delta f\left(r\right)=\frac{1}{2\pi}\left(\Omega_{\rm uni}+\delta\Omega\left(r\right)\right)\quad\hbox{with}\quad \delta\Omega\left(r\right)\!<\!\!<\!\Omega_{\rm uni}.
\end{equation}
In this case, Eq. (\ref{eq:Pg_diff}) becomes
\begin{equation}
2\pi^2\left(n+\alpha_g\right)=\frac{\sqrt{\lambda_{\nu_{\rm uni} km}}}{{\widehat f}_{\rm co}}\int_{r_1}^{r_2}\frac{N\left(r\right)\left[1+{\widehat\varepsilon}\left(r\right)\right]}{r}\mathrm{d}r,
\end{equation}
where $\nu_{\rm uni}=2f_{\rm uni}/{\widehat f}_{\rm co}$, ${\widehat f}_{\rm co}=f_{\rm in}-m f_{\rm uni}$ and ${\widehat\varepsilon}\left(r\right)=m\delta f_{\rm rot}\left(r\right)/{\widehat f}_{\rm co}$.\\

In addition, we can further explore the impact of differential rotation, by rewriting Eq. (\ref{eq:Pg_diff}) using Taylor expansions for $\sqrt{\lambda_{\nu km}\left(r\right)}$ and $(f_{\rm in}-mf_{\rm rot}\left(r\right))^{-1}$ at a radial coordinate $r_s$, where the sensitivity of the g-modes to the stellar structure is dominant. This leads to 
\begin{eqnarray}
\Pi_0\left(n+\alpha_g\right) & \approx & \frac{\sqrt{\lambda_{\nu km,s}}}{f_{\rm co,s}}\left\{1 + \left(\frac{\lambda'_{\nu km,s}}{2\lambda_{\nu km,s}} + \frac{mf'_{\rm rot,s}}{f_{\rm co,s}}\right)\langle r - r_s\rangle\right.\nonumber\\ 
&+& \left(\frac{\lambda''_{\nu km,s}}{2\lambda_{\nu km,s}} - \left(\frac{\lambda'_{\nu km,s}}{2\lambda_{\nu km,s}}\right)^2 + \frac{mf''_{\rm rot,s}}{f_{\rm co,s}} + 2\left(\frac{mf'_{\rm rot,s}}{f_{\rm co,s}}\right)^2\right)\nonumber\\&&\times\left.\vphantom{\left(\frac{\lambda''_{\nu km,s}}{2\lambda_{\nu km,s}} - \left(\frac{\lambda'_{\nu km,s}}{2\lambda_{\nu km,s}}\right)^2 + \frac{mf''_{\rm rot,s}}{f_{\rm co,s}} + 2\left(\frac{mf'_{\rm rot,s}}{f_{\rm co,s}}\right)^2\right)}\langle\left(r - r_s\right)^2\rangle\right\}.
\label{eq:fco_taylor}\end{eqnarray}
Here the quantity $x_s$ denotes the value of the quantity $x$ at the radial coordinate $r_s$, $x' = \frac{\mathrm{d}x}{\mathrm{d}r}$, and
\begin{equation*}
 \langle x\rangle \equiv \left(\int_{r_1}^{r_2}x\frac{N\left(r\right)}{r}\mathrm{d}r\right)\left(\int_{r_1}^{r_2}\frac{N\left(r\right)}{r}\mathrm{d}r\right)^{-1}.
\end{equation*}
The pulsation frequency in the local corotating frame is given by 
\begin{equation}
  f_{\rm co}\left(r\right) = f_{\rm in} - mf_{\rm rot}\left(r\right). \label{eq:rot_frame}
\end{equation}

Because the eigenvalue $\lambda_{\nu km}$ varies smoothly as a function of the spin parameter $\nu$ for a given pulsation mode identification $(k,m)$, the derivatives in the right-hand side of Eq. (\ref{eq:fco_taylor}) can be rewritten using the chain rule. Hereby we consider the functions $\lambda_{km}\left(\nu\right)$, $\nu\left(f_{\rm rot}\right)$ and $f_{\rm rot}\left(r\right)$ for a given pulsation. Taking into account that the pulsation period in the corotating frame $P_{\rm co,s} = f_{\rm co,s}^{-1}$, we find
\begin{equation}
 \Pi_0\left(n+\alpha_g\right) \approx \sqrt{\lambda_{\nu km,s}}\left(P_{\rm co,s} + \sum_{i=1}^3a_iG_i\left(\nu\right)P_{\rm co,s}^{i+1}\right).
\label{eq:Pco_model}
\end{equation}
The coefficients $a_i$ depend on the rotation profile, and are the same for all pulsations. The functions $G_i\left(\nu\right)$ depend on the pulsation mode identification, and describe how the influence of the differential rotation varies across the period spacing patterns. The full expressions for $a_i$ and $G_i\left(\nu\right)$ are given in Appendix \ref{Appendix:coeff}. Because the functions $G_i\left(\nu\right)$ are continuous and smooth, we can conclude that differential rotation in the near-core regions of a g-mode pulsator also introduces an additional continuous change in the slope of the period spacing patterns.

\subsection{MESA \& GYRE models}
\label{subsubsec:illustration}

\begin{figure*}[ht]
\includegraphics[width=\textwidth]{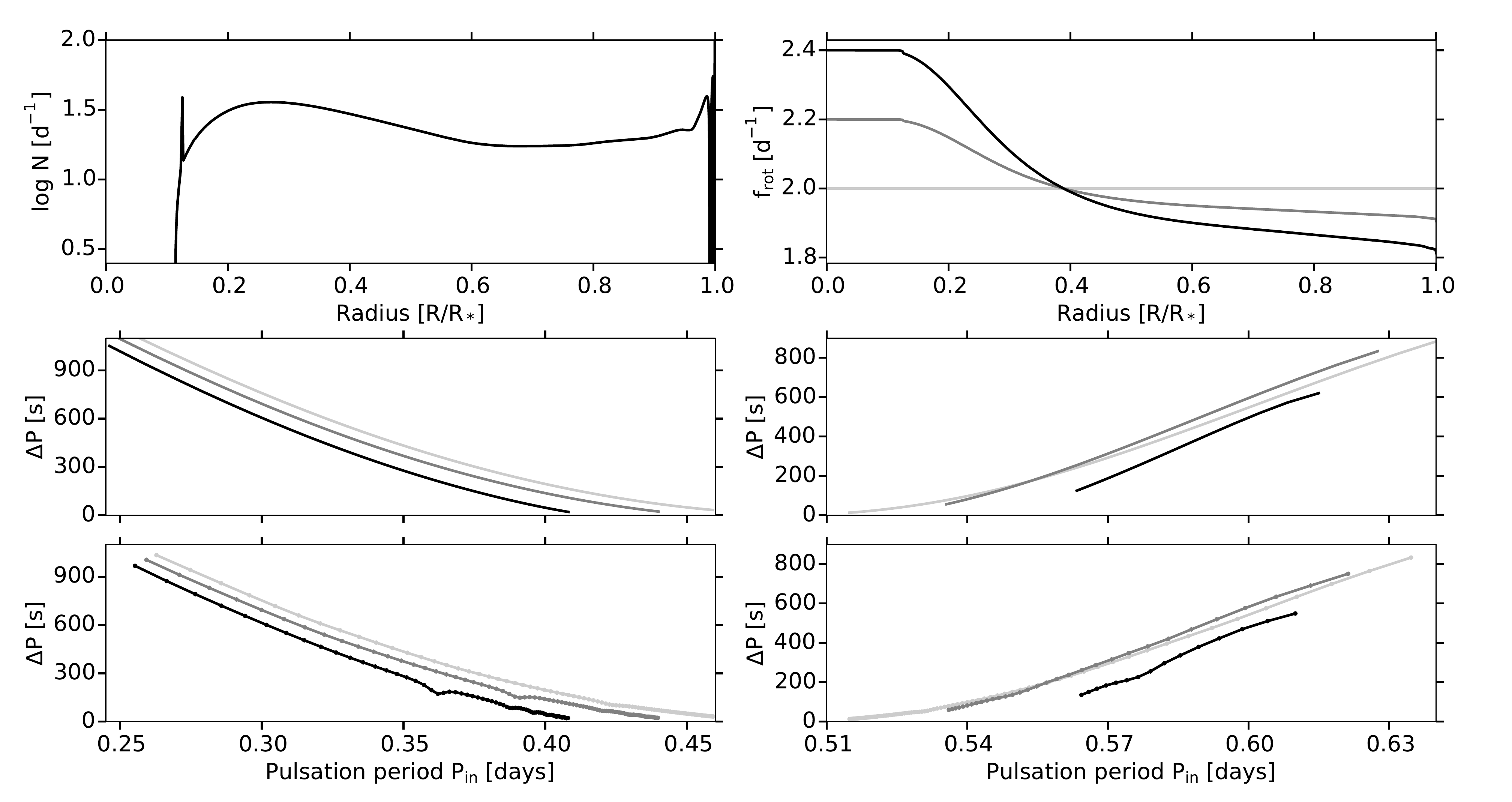}
\caption{\label{fig:mesasim} The $1.7\,M_\odot$ stellar model at $X_c = 0.7$ with $f_{\rm rot,s} = 2\,\rm d^{-1}$ and $\delta f = 0.0$ (light grey), 0.1 ({grey}), and 0.2 (black). {\em Top left:} the Brunt-V\"ais\"al\"a frequency profile. {\em Top right:} The considered stellar rotation profiles. {\em Left:} The corresponding gravito-inertial mode pulsations with $(k,m) = (+0,+1)$, computed from Eq.\,(\ref{eq:Pg_diff}) ({\em middle}) and with GYRE v5.1 ({\em bottom}). {\em Right:} The corresponding gravito-inertial mode pulsations with $(k,m) = (-2,-1)$, computed from Eq.\,(\ref{eq:Pg_diff}) ({\em middle}) and with GYRE v5.1 ({\em bottom}).}
\end{figure*}

 \begin{figure*}[ht]
 \includegraphics[width=\textwidth]{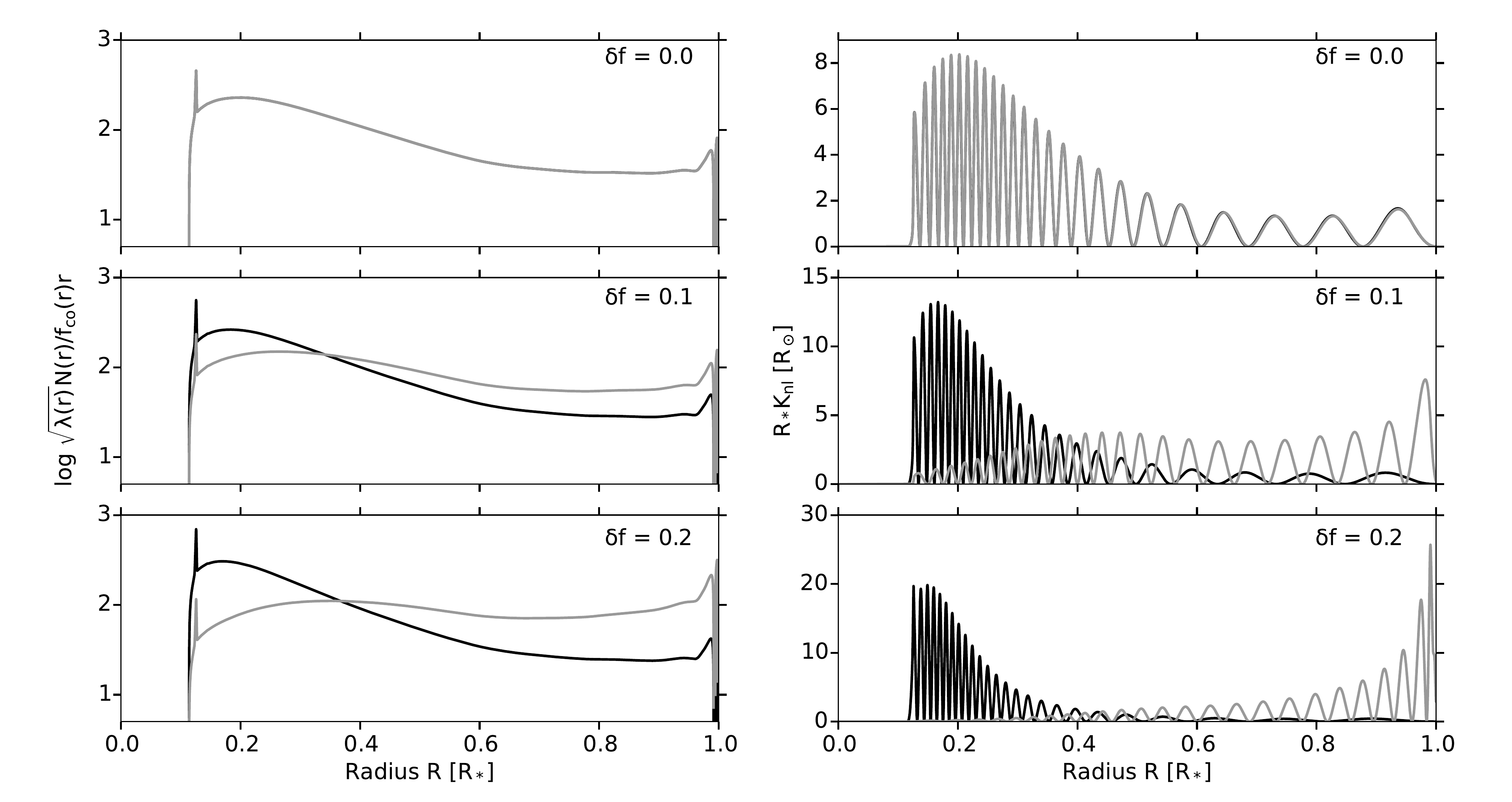}
\caption{\label{fig:kernelsim} The pulsation mode cavities of the gravito-inertial mode pulsations with $(n,k,m) = (-25,+0,+1)$ (black) and $(n,k,m) = (-25,-2,-1)$ (grey) for the $1.7\,M_\odot$ stellar model at $X_c = 0.7$ with $f_{\rm rot,s} = 2\,\rm d^{-1}$, shown in Fig.\,\ref{fig:mesasim}. Both the integrand of Eq. (\ref{eq:Pg_diff}) ({\em left}) and the rotation kernel $K_{nl}$ (computed following Eq. (3.356) in \citet{Aerts2010}; {\em right}) are shown for differential rotation rates $\delta f = 0.0$ ({\em top}), 0.1 ({\em middle}), and 0.2 ({\em bottom}).}
\end{figure*}

We illustrate the theoretical results from Section \ref{subsec:theor} by computing pulsation periods for stellar evolution models including a differential rotation profile. The input models are computed with the one-dimensional stellar evolution code MESA v10108 \citep{Paxton2011,Paxton2013,Paxton2015,Paxton2018}, for a non-rotating 1.7\,$M_\odot$ star with solar metallicity, assuming the chemical mixture described by \citet{Asplund2009} and using OPAL opacity tables \citep{Rogers2002}. Convection is treated using the mixing length theory, with  an $\alpha_{\rm MLT}$ value of 1.8, exponential convective core overshooting of $0.015\,H_p$, and extra diffusive mixing in the radiative region of $1\,\rm cm^2\,s^{-1}$. These values were chosen to match the ones used in studies of g-mode pulsators observed with space photometry \citep[e.g.,][]{Moravveji2015,Moravveji2016,SchmidAerts2016,Zwintz2017}. The differential rotation profile is added a posteriori to the model. It is defined as 
\begin{equation}
 f_{\rm rot}\left(r\right) = f_{\rm rot,s}\left[1 + \delta f\left(\int_r^{r_s}\frac{N^2\left(r\right)}{r}\mathrm{d}r\right)\left(\int_0^{r_s}\frac{N^2\left(r\right)}{r}\mathrm{d}r\right)^{-1}\right],\label{eq:rotprof}
\end{equation}
where $r_s = \langle r\rangle$, and $f_{\rm rot,s}$ and $\delta f$ are free parameters, providing the rotation frequency at radius $r_s$ and the relative difference in rotation frequency between $r = 0$ and $r = r_s$, respectively. This prescription is based on the differential rotation profile derived by \citet{Rieutord2006} and \citet{Hypolite2014} for the thermal wind balance in rapidly rotating stars, taking into account the effects of rotation and stable stratification in the Boussinesq approximation. 

We evaluated the asymptotic relation in Eq.\,(\ref{eq:Pg_diff}) for different theoretical stellar models, and compared our results with full numerical pulsation frequencies computed for the same input models. For the evaluation of Eq.\,(\ref{eq:Pg_diff}), we relied on the numerical Brunt-V\"ais\"al\"a frequency profiles. The pulsations of the stellar models were calculated at multiple points during the main-sequence evolution (with a core mass hydrogen fraction $X_c$ of 0.7, 0.5, 0.3, and 0.1, respectively), for several rotation rates ($f_{\rm rot,s} = 0.1\,{\rm d}^{-1}, 1\,{\rm d}^{-1}$, $1.5\,{\rm d}^{-1}$, and $2\,{\rm d}^{-1}$) with varying degrees of differential rotation. The numerical pulsation frequencies of the stellar models were computed using the adiabatic version of the stellar pulsation code GYRE v5.1 \citep{Townsend2013,Townsend2018}, since stability or excitation rate computations \citep[see, e.g.,][]{Bouabid2013} are beyond the scope of the present work. This version of the GYRE code adopts the TAR to model gravito-inertial and purely inertial modes. The implementation of the TAR in GYRE, which was undertaken independently of the present work, is similar to the approach described by \citet{Townsend2005}, with the allowance for differential rotation as derived by \citet{Mathis2009}.  

 In most cases, our test results were qualitatively similar. A representative case is shown in Fig. \ref{fig:mesasim}. While the asymptotic relation in Eq.\,(\ref{eq:Pg_diff}) does not account for mode trapping, it describes the changes in the mean pulsation period and the slope of the period spacing patterns, caused by differential rotation. The results from Eq.\,(\ref{eq:Pg_diff}) and those obtained numerically with the GYRE code are in agreement. 
 
 The impact of the differential rotation on the pulsation modes depends on the pulsation mode identification. For retrograde gravito-inertial modes, such as r-modes with $(k,m) = (-2,-1)$, the local pulsation frequency in the co-rotating frame is lower when the local rotation frequency is lower, and vice versa, as we can see from Eq. (\ref{eq:rot_frame}). For prograde modes, the local pulsation frequency in the co-rotating frame decreases when the local rotation frequency increases. The relative contribution of the Coriolis force to the total restoring force increases when the local pulsation frequency in the co-rotating reference frame decreases. Hence, retrograde (prograde) modes become more sensitive to the slower (faster) rotating layers in a differentially rotating star. Thus, if the stellar core rotates sufficiently faster than the outer layers, as illustrated in Fig.\ref{fig:mesasim}, the retrograde r-modes are more sensitive to the outer layers of the star, whereas the prograde gravito-inertial modes are more sensitive to the near-core regions, as illustrated in Fig.\,\ref{fig:kernelsim}. Finally, as we can see in Fig.\,\ref{fig:mesasim}, changes in the rotation profile cause the dips in the period spacing patterns to shift in period. This is also indicative of changes in the pulsation mode cavities. 
 
 Other illustrative examples of the impact of differential rotation on g-mode period spacing patterns can be found in the Appendix, in Figs. \ref{fig:yanaisim} and \ref{fig:retrogsim}.

\subsection{Revisiting the $\gamma$\,Dor sample analysis}
\label{subsec:sample}

 \begin{table}[h] 
 \caption{\label{tab:rigidsim} The parameter values of the simulations shown for KIC\,11721304 and KIC\,12066947 in Figs.\,\ref{fig:117mod} and \ref{fig:120mod}, compared with the parameter values of the best model assuming rigid rotation.}
\centering
\begin{tabular}{llllll} 
\hline\hline
\multicolumn{1}{c}{$\Pi_{\rm 0,sim}$} & \multicolumn{1}{c}{$f_{\rm rot,s,sim}$} & \multicolumn{1}{c}{$\delta f_{\rm sim}$} & \multicolumn{1}{c}{$\Pi_{\rm 0,model}$} & \multicolumn{1}{c}{$f_{\rm rot,i,model}$} & \multicolumn{1}{c}{$\chi^2_{\rm red}$} \\
\multicolumn{1}{c}{$[s]$} & \multicolumn{1}{c}{[$\rm d^{-1}$]} & & \multicolumn{1}{c}{[$s$]} & \multicolumn{1}{c}{[$\rm d^{-1}$]} & \\
\hline\\[-4pt]
 \multicolumn{6}{c}{KIC\,11721304}\\[4pt]
 4328 & 0.458 & 0.0  & $4350(230)$ & $0.459(13)$ & 13.1\\
 4105 & 0.339 & 0.88 & $4800(330)$ & $0.483(16)$ & 22.8\\
 \hline\\[-4pt]
 \multicolumn{6}{c}{KIC\,12066947}\\[4pt]
 4180 & 2.16  & 0.0  & $4180(60)$  & $2.161(4)$ & 155.4\\
 4152 & 2.14  & 0.04 & $4110(50)$  & $2.153(4)$ & 114.8\\
 3965 & 2.04  & 0.15 & $3590(120)$ & $2.093(12)$ & 1052\\
\hline
\end{tabular} 
\end{table}

We now reanalyse the period spacing patterns of $\gamma$\,Dor stars reported by \citet{VanReeth2015}, and consider the possibility of radial differential rotation. These stars have been studied in an ensemble modelling analysis by \citet{VanReeth2016}, assuming rigid rotation. This was successful for 40 of the 50 stars with observed period spacings, and led to the identification of r-mode pulsations in ten stars in the sample. These results were recently confirmed in further analyses by \citet{Saio2018}.

\begin{figure}
 \includegraphics[width=88mm]{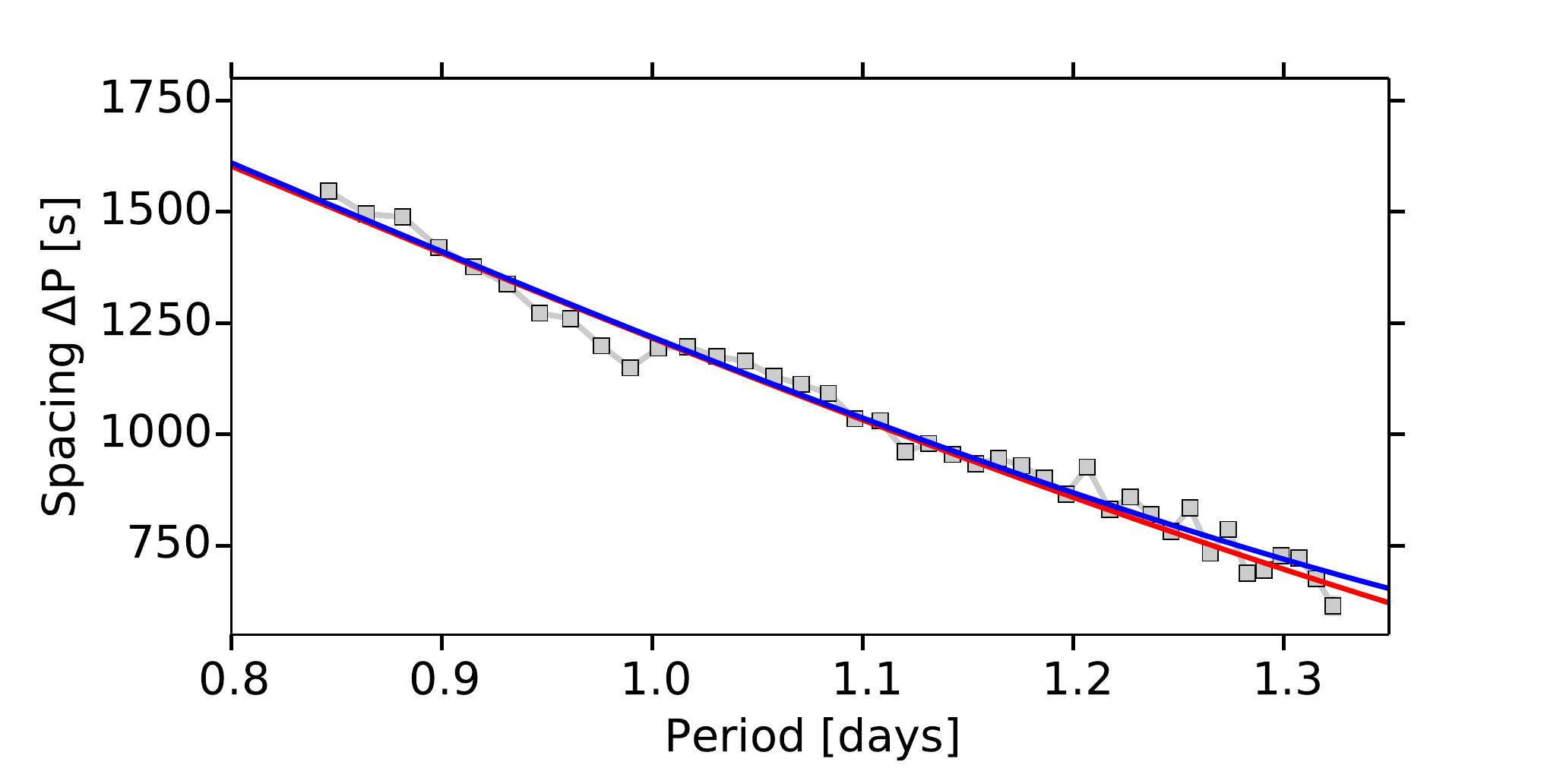}
 \caption{\label{fig:KIC11721304} The observed period spacing pattern of KIC\,11721304 (grey squares) with the best model assuming rigid rotation (red; $f_{\rm rot} = 0.46\pm0.02\,\rm d^{-1}$, $\Pi_0 = 4320\pm230\,s$), and allowing for differential rotation (blue; $f_{\rm rot} = 0.51\,\rm d^{-1}$, $\Pi_0 = 4430\,s$, $a_1 = -0.04$).}
\end{figure}

\begin{figure}
 \includegraphics[width=88mm]{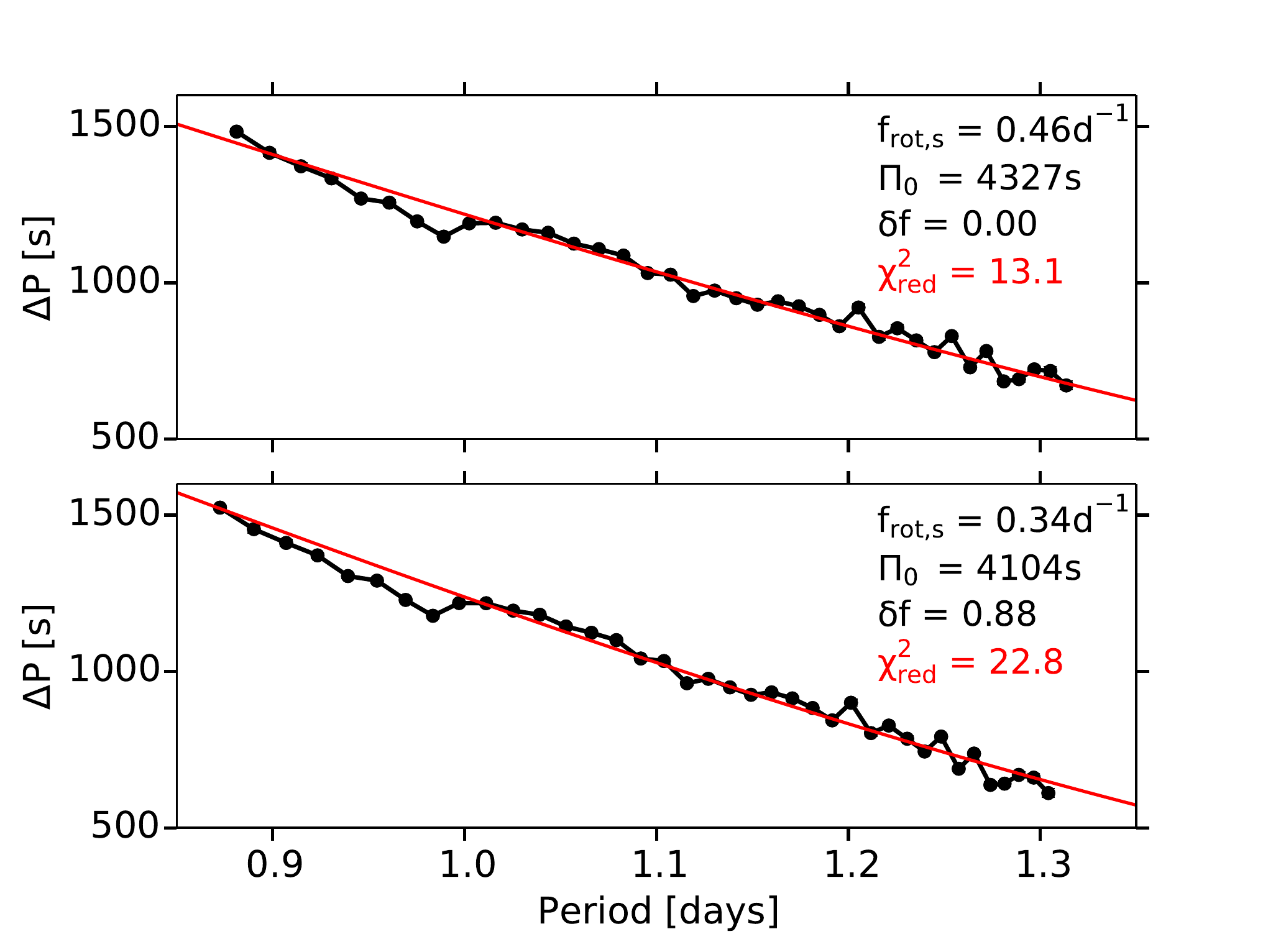}
 \caption{\label{fig:117mod}Simulated period spacing patterns based on the observations for KIC\,11721304 (black dots). The input parameters for the simulations (with non-rigid rotation) are shown in black, while the $\chi^2_{\rm red}$ of the best model with rigid rotation (full red line) is also given.}
\end{figure}

In a first step, we fit the observed period spacing patterns using Eq. (\ref{eq:Pco_model}), whereby we optimise the parameters $\Pi_0$, $f_{\rm rot}$, and $a_i$. 
From this analysis we find that the observed period spacing patterns often do not allow to detect differential rotation.
This is illustrated in Fig.\,\ref{fig:KIC11721304} for the $\gamma$\,Dor star KIC\,11721304, where we show the observed pattern, the best fitting model assuming rigid rotation, and a model that includes differential rotation. The quality of the fit is similar for both types of rotation profiles, but when differential rotation is included, the uncertainties on the derived near-core rotation rate and the asymptotic spacing become larger. Hence, the model assuming rigid rotation is statistically favourable, but we cannot disprove differential rotation. 

To properly establish the extent of this degeneracy, we simulate period spacing patterns for varying degrees of differential rotation, and analyse them using rigidly rotating models. The simulated period of the pulsation with mode identification $(n,k,m)$ in the inertial reference frame, is given by
\begin{equation}
 P_{nkm} = P_{{\rm th},nkm} + \delta P_{nkm}. \label{eq:simul}
\end{equation}
Here, $P_{{\rm th},nkm}$ is the theoretical pulsation period for the mode $(n,k,m)$ in the asymptotic regime, computed with Eq. (\ref{eq:Pg_diff}). 
We used the $1.7M_\odot$ MESA model discussed in Section \ref{subsubsec:illustration} at $X_c = 0.5$ as input, combined with rotation profiles that were constructed following Eq. (\ref{eq:rotprof}) for increasing differential rotation rates ($0\leq\delta f\cdot f_{\rm rot,s} \leq 0.3$). The offsets $\delta P_{nkm}$ in Eq.\,(\ref{eq:simul}) account for the effects of mode trapping and observational errors that were included in our simulated data. The latter were taken to be the residuals after modelling an observed period spacing pattern. 
These simulated period spacing patterns were subsequently fitted with a rigid rotation model, as described by Eq. (\ref{eq:Pg_uni}), and the quality of the fits were evaluated using reduced $\chi^2$-values. 

\begin{figure*}[ht]
 \includegraphics[width=\textwidth]{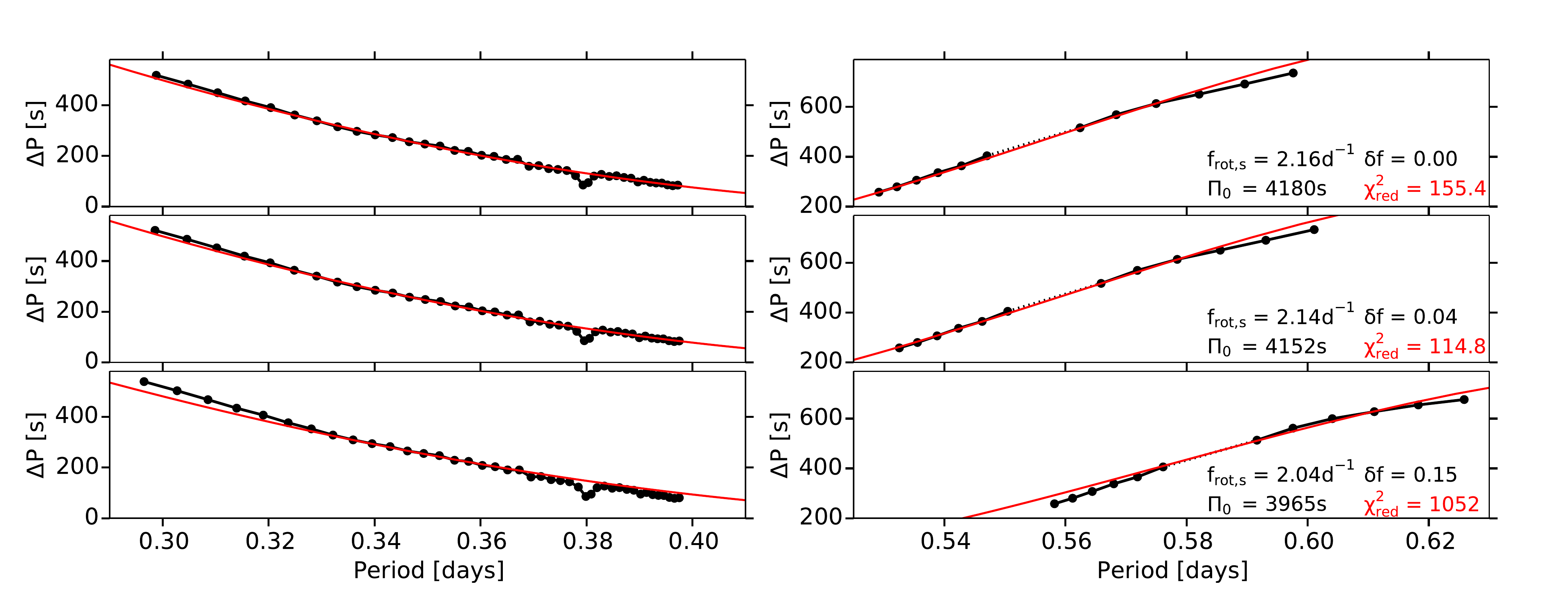}
 \caption{\label{fig:120mod}Simulated period spacing patterns for different rotation profiles {\em (top, middle, bottom)} based on the observations for KIC\,12066947 (black dots). The input parameters for the simulations (with non-rigid rotation) are shown in black on the right, while the $\chi^2_{\rm red}$ of the best model with rigid rotation (full red line) is also given. Both the prograde dipole modes (with $(k,m)$ = (0,1); {\em left}) and the r-modes (with $(k,m)$ = (-2,-1); {\em right}) are shown.}
\end{figure*}

The results of two selected simulations, with offsets $\delta P_{nkm}$ based on the observations of the $\gamma$\,Dor star KIC\,11721304, are shown in Fig.\,\ref{fig:117mod} and listed in Table\,\ref{tab:rigidsim}. As the differential rotation $\delta f$ in the simulated data increases, the quality of the rigidly rotating model fit decreases. However, this is hard to observe. Indeed, differences in the slope of the simulated pattern caused by the differential rotation are accounted for by slight changes in the near-core rotation rate $f_{\rm rot,i}$ and asymptotic spacing $\Pi_0$ of the rigidly rotating model. The changes in the model value of $\Pi_0$ are related to the interaction between the differential rotation profile and the Brunt-V\"ais\"al\"a frequency profile, as we derived from Eq.\,(\ref{eq:Pg_diff}) and illustrated in Fig.\ref{fig:kernelsim}.
For all simulated differential rotation profiles, the values for $f_{\rm rot,i}$ and $\Pi_0$ of the best model are as typically detected for a $\gamma$\,Dor star. Any remaining differences between the differentially rotating simulation $P_{{\rm th},nkm}$ and the best fitting model are negligible compared to the offsets $\delta P_{nkm}$. As a result, we cannot properly detect the differential rotation in these simulated data, even for $\delta f\approx 0.9$.

%

The degeneracy between rigid and differential rotation is lifted when multiple period spacing patterns are observed for the same star. This is illustrated in Fig.\,\ref{fig:120mod}. Here we have simulations of a prograde dipole mode series and a retrograde r-mode series, with offsets $\delta P_{nkm}$ based on the observations of the $\gamma$\,Dor star KIC\,12066947. The parameter values of the simulated patterns and the best (rigidly rotating) models are again listed in Table~\ref{tab:rigidsim}. Because a differential rotation profile has a different effect on both patterns (as discussed in Section \ref{subsubsec:illustration}), its presence is revealed when models assuming uniform rotation are fitted simultaneously to the data. We can see that for the simulations in the bottom panels of Fig.\,\ref{fig:120mod}, with $\delta f = 0.15$, the rigidly rotating model fails to match the slope of the patterns. Since we did not observe such significant shortcomings when modelling the real observed period spacing patterns for this star, we estimate that its differential rotation rate is $\delta f \lesssim 0.15$. Similar model-dependent upper limits for $\delta f$ have been estimated for the other stars with detected r-mode patterns, and are listed in Table\,\ref{tab:surfcore}.

In addition, the simulations illustrate another property when dealing with non-rigid rotation. The uniformly rotating model provides a better fit to the simulated data with a differential rotation rate $\delta f = 0.04$, than to the simulated data for a rigid rotator ($\delta f = 0.0$), as evidenced by their respective $\chi^2_{\rm red}$ values. For $\delta f = 0.04$, the observational signatures of pulsation mode trapping by a chemical gradient and the differential rotation partially cancel each other. Modelling of the patterns in Fig.\,\ref{fig:120mod} individually confirms these results and is in line with the discussion in Section \ref{subsubsec:illustration}. As the differential rotation increases, the prograde gravito-inertial modes become more sensitive to the faster rotating near-core regions, and the r-modes become more sensitive to the slower rotating outer layers of the star. At the differential rotation rate $\delta f = 0.15$, the results from modelling both patterns individually no longer agree within their respective errors. In addition, the r-modes have become sufficiently sensitive to the outer layers of the star that the derived asymptotic spacing $\Pi_0$ from the modelling, as computed in Eq.\,(\ref{eq:pi0}), is no longer reliable.

\subsection{Discovery of Yanai modes}
\label{subsubsec:yanai}
For ten stars in the sample, \citet{VanReeth2016} could not find a unique pulsation mode identification for the detected period spacing patterns. For some, there were multiple mode identifications that yielded equally good results. However, for six of them, no suitable interpretation could be found at all. Interestingly, all of these stars have a few characteristics in common. They have low to moderate values for the spectroscopic rotation velocity $v\sin\,i$ (see Table \ref{tab:yanai}). For each of them \citet{VanReeth2015} detected a single period spacing pattern with an upward slope, i.e. of retrograde pulsation modes, with period spacing values typically ranging from a few hundred to a few thousand seconds. Qualitatively, these characteristics agree with what is expected for Yanai modes, a type of purely inertial pulsation modes with $k=-1$, otherwise known as odd r-modes \citep{Townsend2003c,Saio2018}. The observed period spacing values are far larger than the spacings of r-modes with $k=-2$, while the spacings of normal retrograde g-modes with $k\geq0$ are not expected to be much smaller than the asymptotic spacing $\Pi_0/\sqrt{l(l+1)}$. However, \citet{VanReeth2016} could not compute a satisfactory theoretical model for Yanai modes, and were limited to the asymptotic expressions given by \citet{Townsend2003c}. 

\begin{figure}
 \includegraphics[width=88mm]{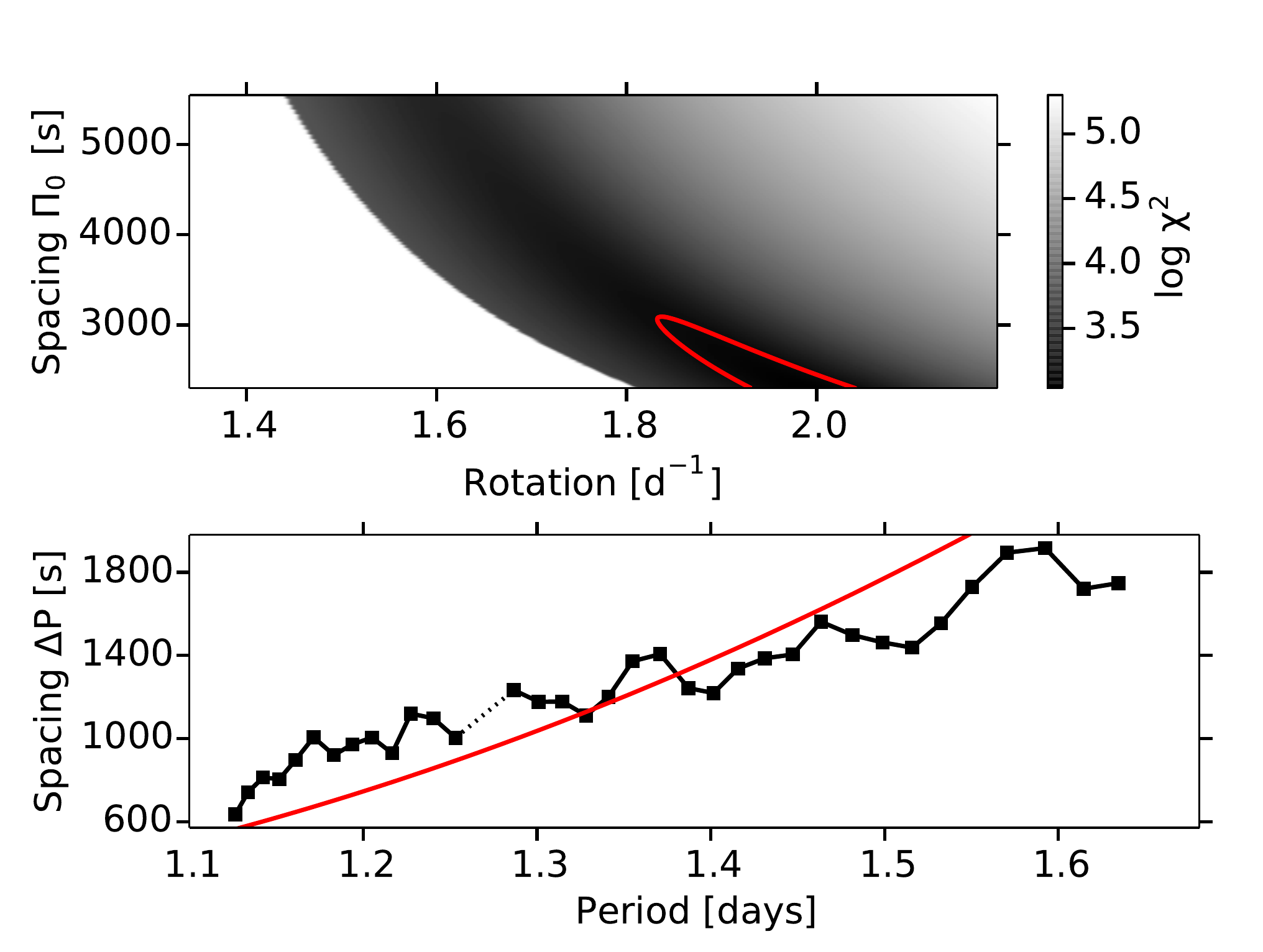}
 \caption{\label{fig:11668783} {\em Top:} The $\log\,\chi^2$-distribution for the parameters of the model pattern evaluated for KIC\,11668783. The red contour indicates the boundary of the 1$\sigma$ confidence interval. {\em Bottom:} The observed period spacing pattern of KIC\,11668783 (black squares) reported by \citet{VanReeth2015}, with the best model (red line; $f_{\rm rot} = 1.99^{+0.05}_{-0.15}\,\rm d^{-1}$, $\Pi_0 = 2300^{+790}_{-0}$\,s).}
\end{figure}

In this work, we are able to compute adequate theoretical models for Yanai modes, using GYRE v5.1. The period spacing pattern of the $\gamma$\,Dor star KIC\,11668783 is shown in Fig.\,\ref{fig:11668783} as an example. The best fitting model, included in the figure, corresponds to Yanai modes with $(k,m) = (-1,-1)$ in a rigidly rotating star. While we can obtain pulsation periods in the observed range, the slope of the model pattern differs from the slope of the observations, and the asymptotic spacing $\Pi_0$ and near-core rotation frequency $f_{\rm rot}$ of our best model have unrealistic values, i.e., $2300^{+790}_{-0}$\,s and $1.99^{+0.05}_{-0.15}\,\rm d^{-1}$ respectively. The inclusion of radial differential rotation in the models was insufficient to resolve this issue. 

\begin{figure}
 \includegraphics[width=88mm]{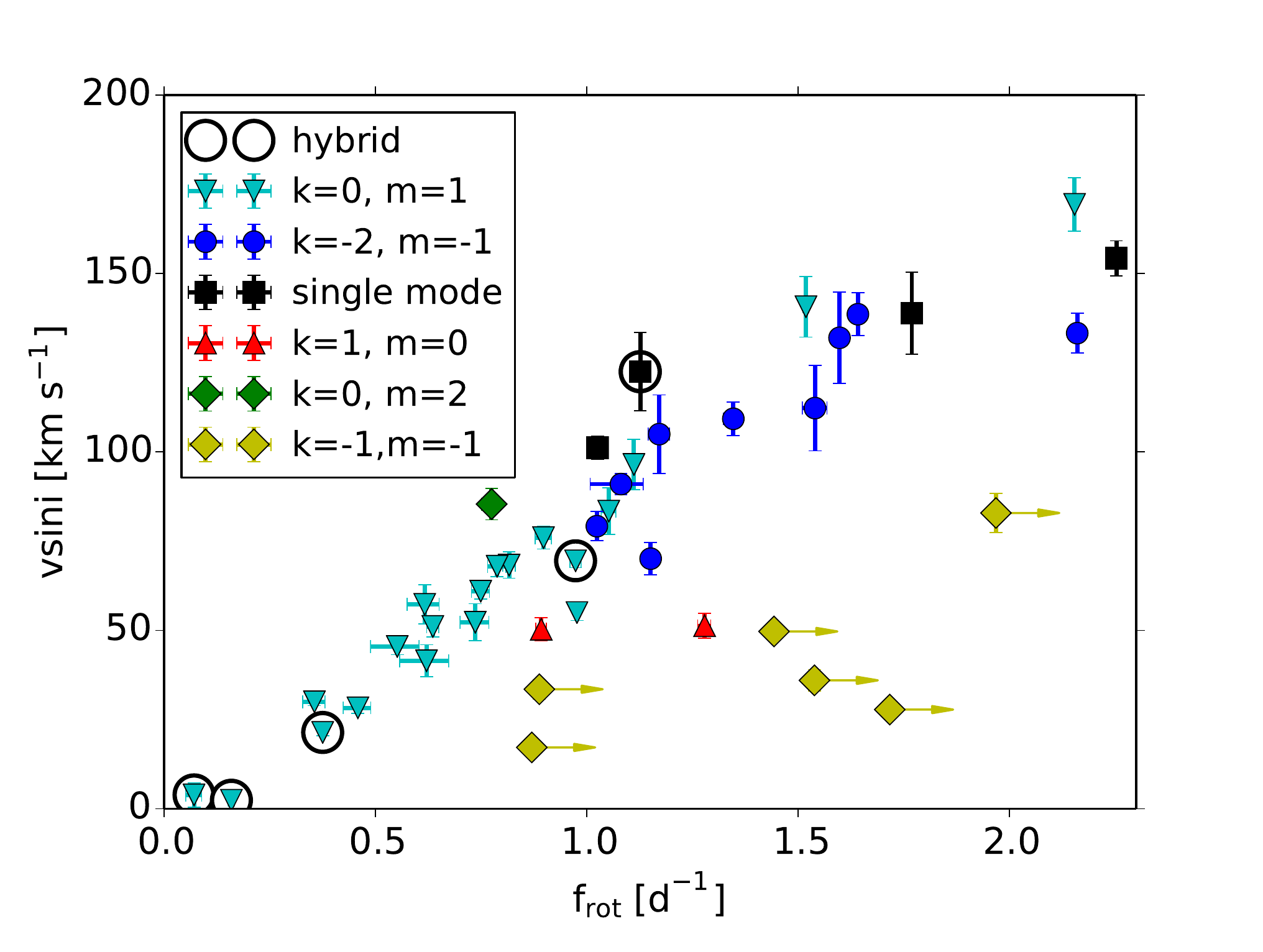}
 \caption{\label{fig:vsiniyanai} The spectroscopic rotation velocities of the single stars in our sample, as a function of the derived near-core rotation rate, assuming uniform rotation as the statistically favoured model. The symbol and colour indicate the derived pulsation mode identification. Contrary to the other stars in this figure, the stars with Yanai modes (with $(k,m) = (-1,-1)$) have not been successfully modelled. For these stars, the plotted rotation frequency is a lower limit, assuming the proposed mode identification is correct. The arrows indicate how the graph changes when $f_{\rm rot}$ is higher. This figure is an update of Fig.\,14 by \citet{VanReeth2016}.}
\end{figure}

\begin{figure}
 \includegraphics[width=88mm]{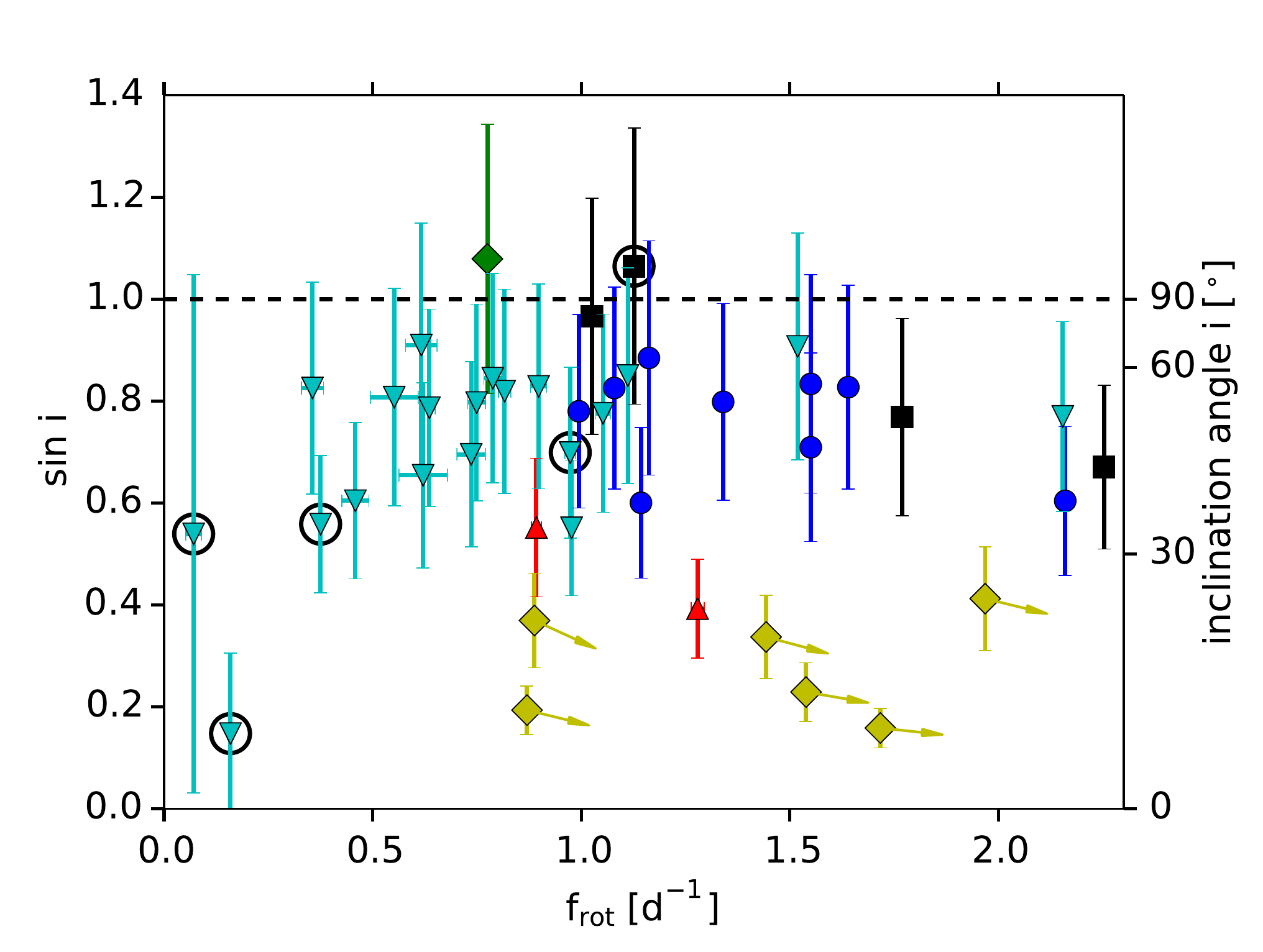}
 \caption{\label{fig:inclyanai} The inclination angle $i$ of the observed stars, derived from the information shown in Fig. \ref{fig:vsiniyanai}, assuming the stars are spherically symmetric with a stellar radius $R_* = 2.0\pm0.5\,R_\odot$, i.e., typical for a $\gamma$\,Dor star. The symbols and colours are the same as in Fig. \ref{fig:vsiniyanai}. For the stars with Yanai modes (with $(k,m) = (-1,-1)$), the plotted rotation frequency is again a lower limit. The arrows indicate how the graph changes when $f_{\rm rot}$ is higher.}
\end{figure}

However, the observations are qualitatively consistent with Yanai modes with $(k,m)=(-1,-1)$. If we assume that the star is a solid body rotator, we can use the mode identification to derive a lower limit for the stellar rotation frequency $f_{\rm rot}$. As explained in detail by \citet{Saio2018}, an r-mode pulsation frequency in the inertial reference frame $f_{\rm in,s} < |m|f_{\rm rot}$, so that for $m=-1$, we have $\left(|m|P_{\rm min}\right)^{-1} < f_{\rm rot}$, where $P_{\rm min}$ is the shortest pulsation period of the detected period spacing pattern. Combined with the spectroscopic values for the projected surface rotation velocity $v\sin i$, we find that these are moderate to fast rotating stars seen close to pole-on. These results, shown in Figs. \ref{fig:vsiniyanai} and \ref{fig:inclyanai}, are generally consistent with the theoretical expectations for Yanai modes \citep{Saio2018}.

 \begin{table}[h] \small
 \caption{\label{tab:yanai} The six stars with single retrograde period spacing patterns, that we suspect correspond to Yanai modes. The listed parameters include the spectroscopic rotation velocity, the dominant pulsation frequency, and the minimum and maximum pulsation periods for the pulsations in the detected period spacing patterns.}
\centering
\begin{tabular}{rllll}
\hline\hline
 KIC & $v\sin i$ & $f_{\rm dom}$ & $P_{\rm min}$  & $P_{\rm max}$\\
 & [km\,$\rm s^{-1}$] & [$\rm d^{-1}$] &  [days] & [days]\\
\hline
 4757184  & 36(3) & 1.26725(2)  & 0.64993(4) & 1.00612(7)\\
 5350598  & 28(2) & 2.158490(4) & 0.58249(2) & 1.36381(8)\\
 6185513  & 83(6) & 2.245965(7) & 0.50812(2) & 1.3940(1) \\
 6425437  & 50(3) & 0.89681(1)  & 0.69305(6) & 1.2640(2) \\
 7867348  & 17(1) & 1.16998(1)  & 1.14962(2) & 1.8273(2) \\
 11668783 & 34(3) & 0.627995(7) & 1.12644(4) & 1.65470(8)\\
\hline
\end{tabular} 
\end{table}

There are differences between the models and the observations for these six stars. 
These stars are fast rotators, seen close to pole-on as indicated by the low $v\sin i$ values. Hence, the traditional approximation may not be applicable to Yanai modes in these stars. The behaviour of Yanai modes becomes similar to that of retrograde gravity-modes at high spin parameter values \citep{Townsend2003c,Saio2018}. Recently, \citet{Ouazzani2017} compared the traditional approximation with a general two-dimensional treatment of stellar pulsations and found that it is reasonably appropriate for g-mode period spacing patterns with various mode identifications, except for retrograde g-modes in moderate to fast rotating stars. In this case, general two-dimensional models are needed. 

\section{Rotational modulation}
\label{sec:rotmod}
A common approach to observationally detect or disprove differential rotation is to determine the ratio between the rotation frequency in the near-core regions and close to or at the stellar surface. While g-mode pulsations are the only reliable way to probe the stellar properties near the convective core, the rotation rate in the outer stellar envelope can be determined by analysing p-mode pulsations \citep[e.g.,][]{Kurtz2014,Saio2015,SchmidAerts2016} and/or rotational spot modulation \citep[e.g.,][]{Degroote2011, EscorzaSantos2016, Saio2018} in the light curve, if either are present. In this section, we focus on detecting rotational modulation and distinguishing it from g-mode pulsations. To avoid confusion, the rotation frequency in the deep stellar interior and in the outer layers are denoted by $f_{\rm rot,i}$ and $f_{\rm rot,o}$, respectively.

When we look for observational signatures of rotational spot modulation in the Fourier spectrum of space-based photometry of a star, we expect to find low-frequency peaks for which the $1^{\rm st}$ harmonic frequency is present as well. These frequencies are often part of a  group of two or more closely spaced frequencies which are not always resolved. Such signal is common for low-mass stars with convective envelopes, but has been observed for A- and B-type stars as well \citep[e.g.,][]{Degroote2011, EscorzaSantos2016, Bowman2017, Saio2018}. 

For non-radial g-mode pulsators, the detection of rotational modulation signal is not straightforward. These pulsation modes are known to have combination frequencies in the observations, and, for main-sequence stars, they occur in the same frequency range as the stellar rotation frequency. Furthermore, we may expect similar amplitudes for rotational modulation and g-mode pulsations in photometric data, and both types of signal are frequently non-sinusoidal. However, contrary to g-mode pulsations, which have lifetimes longer than the total observation time of the {\em Kepler} space mission, rotational spot modulation can vary in amplitude on much shorter timescales. In addition, whereas g-mode pulsations have very stable frequencies on long timescales, rotational spot modulation frequencies often exhibit small variations dependent on the latitude of the stellar spots. Consequently, rotational modulation signal ofttimes consists of isolated closely spaced frequency groups, clustered around harmonic frequencies.

Hence, we evaluate the frequencies extracted from the {\em Kepler} photometry by \citet{VanReeth2015} to look for rotational modulation signal. To properly distinguish g-mode pulsations from rotational modulation, we require that
\begin{itemize}
 \item the signal-to-noise ratio $S/N \geq 4$, and the value of the amplitude is at least five times as large as its error, to ensure that the detected signal is significant.
 \item the first harmonic of the selected frequencies is detectable in the data, whereby we use the Rayleigh frequency resolution $f_{\rm res}$ as a limit to find harmonics,
 \item the selected frequency is part of a closely spaced group of frequencies, whereby $|f_i - f_j| < 1.5f_{\rm res}$, 
 \item both the selected frequencies and their first harmonics are located outside of the frequency ranges of the g-mode pulsation frequencies and their combinations,
 \item the selected frequencies are larger than half the derived near-core rotation frequency $f_{\rm rot,i}$ and any detected r-mode frequencies,
 \item the selected frequencies are smaller than one-and-a-half times the derived near-core rotation frequency $f_{\rm rot,i}$ and any detected prograde pulsation mode frequencies.
\end{itemize}

With these criteria, the main limitation is that stellar spots must have a shorter lifetime than the total observation time of the original {\em Kepler} space mission. This implies that we cannot detect rotational modulation in g-mode pulsators with long-lived stellar spots. A second limitation of our approach is that we require the surface rotation rate and the near-core rotation rate to agree within 50\%. This assumption would be unjustified in (very) slowly rotating stars, such as KIC\,8197761 \citep{Sowicka2017}. However, the stars in our sample rotate significantly faster \citep{VanReeth2016}. 

\begin{figure*}
 \includegraphics[width=\textwidth]{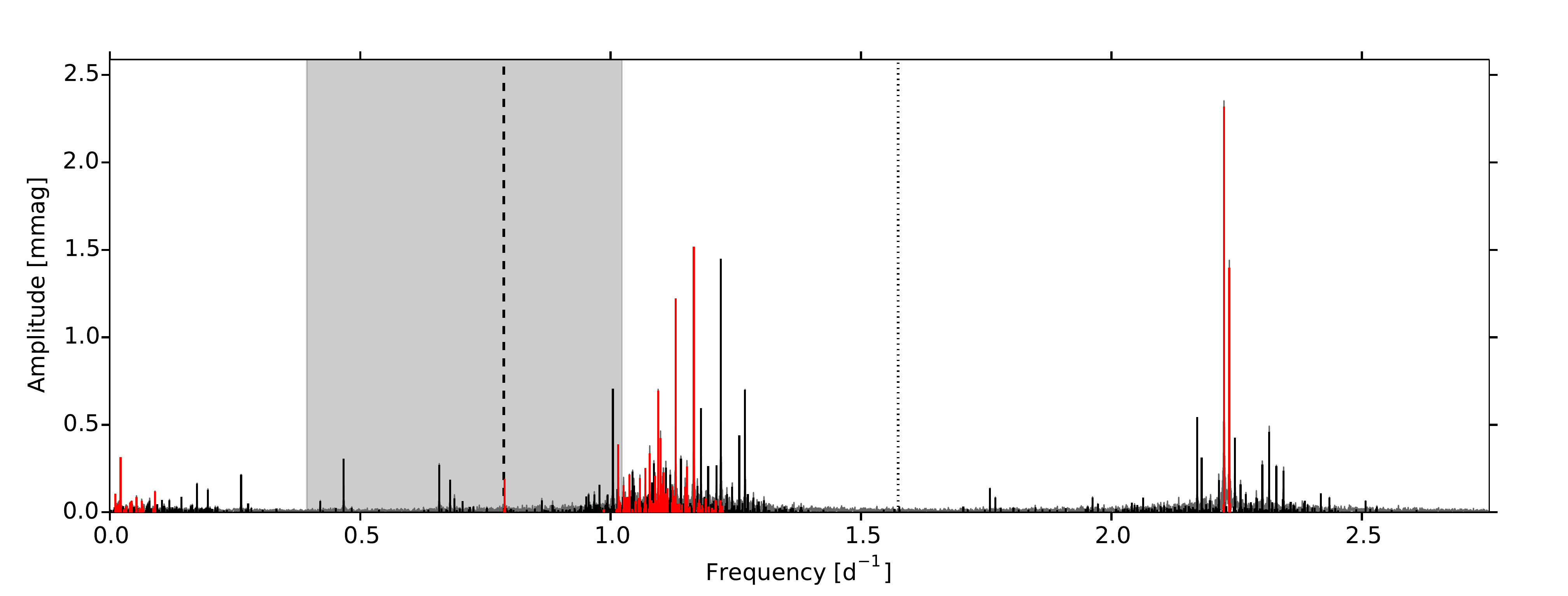}
 \caption{\label{fig:11294808} The Fourier spectrum of KIC\,11294808. Frequencies that fulfil our detection criterion, are indicated in black, whereas frequencies marked in red fulfil the detection criterion, have the first-order harmonic within the extracted frequency list, and are part of a closely spaced group of frequencies. The derived near-core rotation frequency and its harmonic are indicated with the dashed and dotted lines, respectively. Signatures of rotational modulation occur within the shaded grey frequency range.}
\end{figure*}

Our analysis is illustrated in Fig.\,\ref{fig:11294808} for the $\gamma$\,Dor star KIC\,11294808. The frequencies marked in red are the only peaks that fulfil the first three criteria, and any frequencies within the grey area fulfil the last two criteria. There is only one peak, at $0.788\pm0.001\,\rm d^{-1}$, that satisfies all of our criteria, and is reasonably isolated within the frequency spectrum. Thus, this signal is caused by rotational modulation, and its frequency is the surface rotation rate $f_{\rm rot,o}$. Consequently, $f_{\rm rot,o}$ and $f_{\rm rot,i}$ are equal within $1\sigma$, indicating that KIC\,11294808 is a rigid rotator. 

\begin{figure*}
 \includegraphics[width=\textwidth]{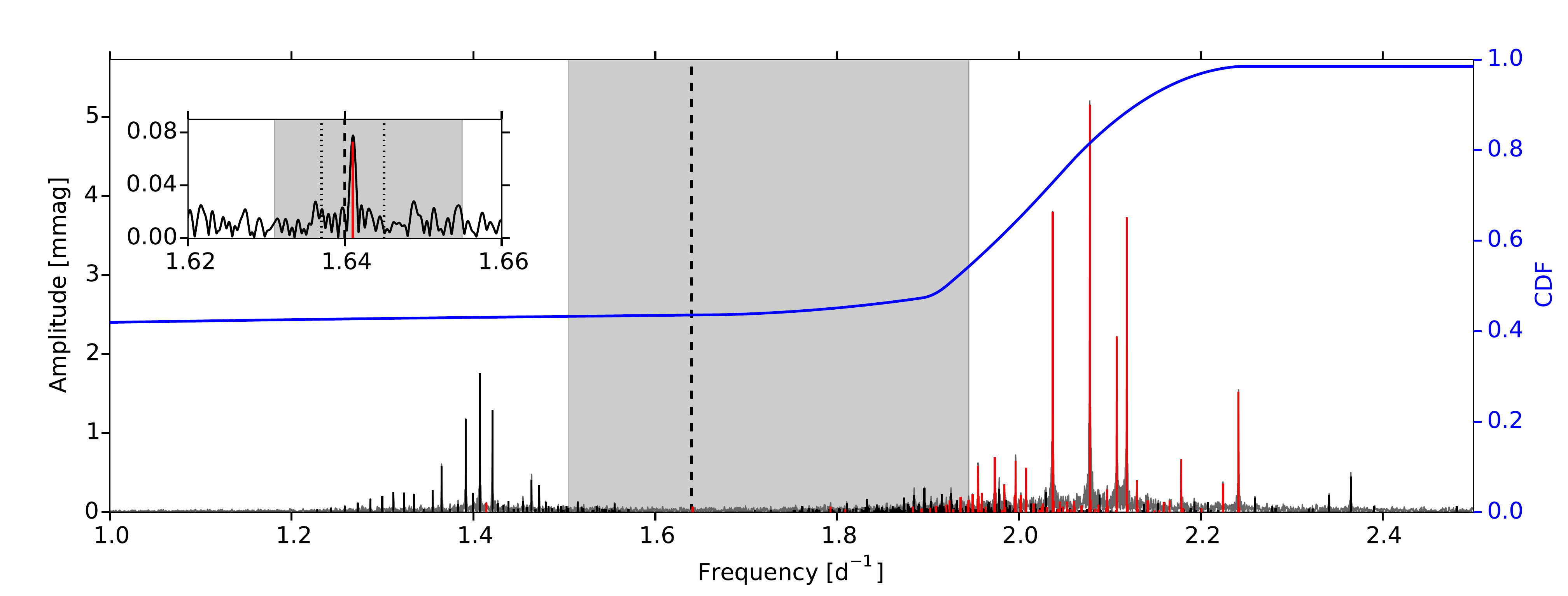}
 \caption{\label{fig:8375138}  The Fourier spectrum of KIC\,8375138. Frequencies that fulfil our detection criterion, are indicated in black, whereas frequencies marked in red fulfil the detection criterion, and have the first-order harmonic within the extracted frequency list. The derived near-core rotation frequency and its harmonic are indicated with the dashed and dotted lines, respectively, while the cumulative distribution function of the red frequency peaks is shown by the blue curve. Signatures of rotational modulation occur within the shaded grey frequency range. {\em Inset:} a close-up of the Fourier spectrum around the near-core rotation rate $f_{\rm rot,i}$ (dashed line). The $1\sigma$ (dotted lines) and $3\sigma$ (grey area) uncertainties on $f_{\rm rot,i}$ are indicated. The ten highest-amplitude pulsation frequencies have been prewhitened for clarity.}
\end{figure*}

To avoid false detections, our employed criteria for the detection of rotational modulation are rather strict. However, when one of the criteria is not met for a star, it can still exhibit rotational modulation. As an example, we show the Fourier spectrum of the $\gamma$\,Dor star KIC\,8375138 in Fig.\,\ref{fig:8375138}. There is a small-amplitude peak in the spectrum at a frequency of $1.641\pm0.001\,\rm d^{-1}$ that complies with all but one of our criteria, i.e., it is not part of a closely spaced group of frequencies, but an isolated frequency. Nevertheless, this is very likely also rotational modulation signal. Thanks to the detection of both prograde dipole modes and r-modes in this star, the near-core rotation rate $f_{\rm rot,i}$ is exceptionally well-constrained, and the detected frequency agrees with $f_{\rm rot,i}$ within a fraction of $1\sigma$. The probability that this is a pulsation frequency, is very small. As shown in Fig.\,\ref{fig:8375138}, we numerically calculated the cumulative distribution function of the detected significant frequencies of this star, that have a first-order harmonic. By drawing random frequency values from the computed distribution function, taking into account the total number of such detected frequencies, we find that there is a 2.1\% chance that a significant pulsation frequency with a first-order harmonic in the data, is located within $1\sigma$ of $f_{\rm rot,i}$. The probability of detecting a random significant pulsation frequency with a first-order harmonic within $3\sigma$ of $f_{\rm rot,i}$, is 7.1\%.

All stars with derived values for the surface rotation rate $f_{\rm rot,o}$ are listed in Table \ref{tab:surfcore}. 

\section{Discussion and Conclusions}
\label{sec:conclusions}
 \begin{table}[h]
 \caption{\label{tab:surfcore} An overview of the stars with derived constraints on differential rotation. The listed values are the near-core rotation rate $f_{\rm rot,i}$, estimated model-dependent upper limits on the near-core differential rotation $\delta f$, the surface rotation rate $f_{\rm rot,o}$, and the ratio $f_{\rm rot,o}/f_{\rm rot,i}$.} 
\centering
\begin{tabular}{lllll}
\hline\hline
KIC  & $f_{\rm rot,i}$ [$\rm d^{-1}$] & $\delta f$ & $f_{\rm rot,o}$ [$\rm d^{-1}$] & $f_{\rm rot,o}/f_{\rm rot,i}$\\ 
\hline
2710594 & 0.995(8) & 0.13 & - & - \\
3448365 & 1.079(3) & 0.11 & - & - \\
5114382 & 1.140(6) & 0.15 & - & - \\
5708550 & 0.82(2) & - & 0.7955(10)   & 0.97(2)\\ 
6468987 & 1.546(6) & 0.13 & - & - \\
7365537 & 2.253(3) & - & 2.242(5)   & 0.995(3)\\ 
7380501 & 0.64(1) & - & 0.6394(10)   & 1.00(2)\\ 
7583663 & 1.162(5) & 0.16 & - & - \\
8364249 & 1.519(8) & - & 1.517(5)   & 0.999(6)\\ 
8375138 & 1.640(5) & 0.12 & 1.641(1)   & 1.001(3)\\ 
9210943 & 1.705(11) & 0.07 & 1.702(15) & 0.998(7)\\ 
9480469 & 1.541(4) & 0.09 & - & - \\
11294808 & 0.77(2) & - &  0.788(1) & 1.02(3)   \\ 
11456474 & 1.05(2) & - & 1.0495(10) & 1.00(2) \\ 
11907454 & 1.341(3) & 0.12 & - & - \\
12066947 & 2.159(4) & 0.15 & - & - \\
\hline
\end{tabular} 
\end{table}

The aim of this work was to provide observational constraints of differential rotation in a sample of $\gamma$\,Dor stars. We approached this issue using two separate methodologies.

First, we considered the impact of differential rotation on the observational properties of g-mode period spacing patterns. We use a theoretical formalism that allows us to study the properties of low-frequency gravito-inertial modes propagating in differentially rotating stars developed in \cite{Mathis2009} and an extended observation sample of $\gamma$ Dor stars. In a uniformly rotating star, the slope of an observed period spacing pattern provides information on the stellar rotation rate, while the non-uniform structure of the pattern is a signature of the pulsation mode trapping caused by chemical gradients and discontinuities in the near-core regions. Differential rotation modifies these diagnostics. It changes the sensitivity of pulsation modes to different depths within the interior structure, which in turn leads to changes in the pulsation mode trapping. This translates into changes of the morphology of the period spacing patterns. A high differential rotation rate will also change the slope of the observed period spacing pattern. Here, we developed a new theoretical seismic diagnostic for the case of radial differential rotation.

In the case where only one period spacing pattern is detected for a star, it is difficult to detect the signature of differential rotation in the pattern. Indeed, other physical processes, such as the mode trapping by chemical gradients, can have similar observational characteristics in the pulsation period spectrum. On the other hand, a period spacing pattern of a differentially rotating star can often be fitted by a rigidly rotating model, where the uniform rotation rate and asymptotic period spacing parameters of the model are tweaked to compensate for the offset. In many cases the rigidly rotating model is statistically favourable over a model with differential rotation. We demonstrated this for the $\gamma$\,Dor star KIC\,11721304, and were unable to detect differential rotation rates below $\delta f < 0.9$.

When multiple period spacing patterns have been detected, the period spacing patterns can be compared to infer the presence of, or an upper limit for, differential rotation. We have demonstrated this for the $\gamma$\,Dor star KIC\,12066947, and found $\delta f \lesssim 0.15$. Similar model-dependent upper limits for $\delta f$ have been estimated for our other stars with detected r-mode patterns, and are listed in Table\,\ref{tab:surfcore}.

\begin{figure}
 \includegraphics[width=88mm]{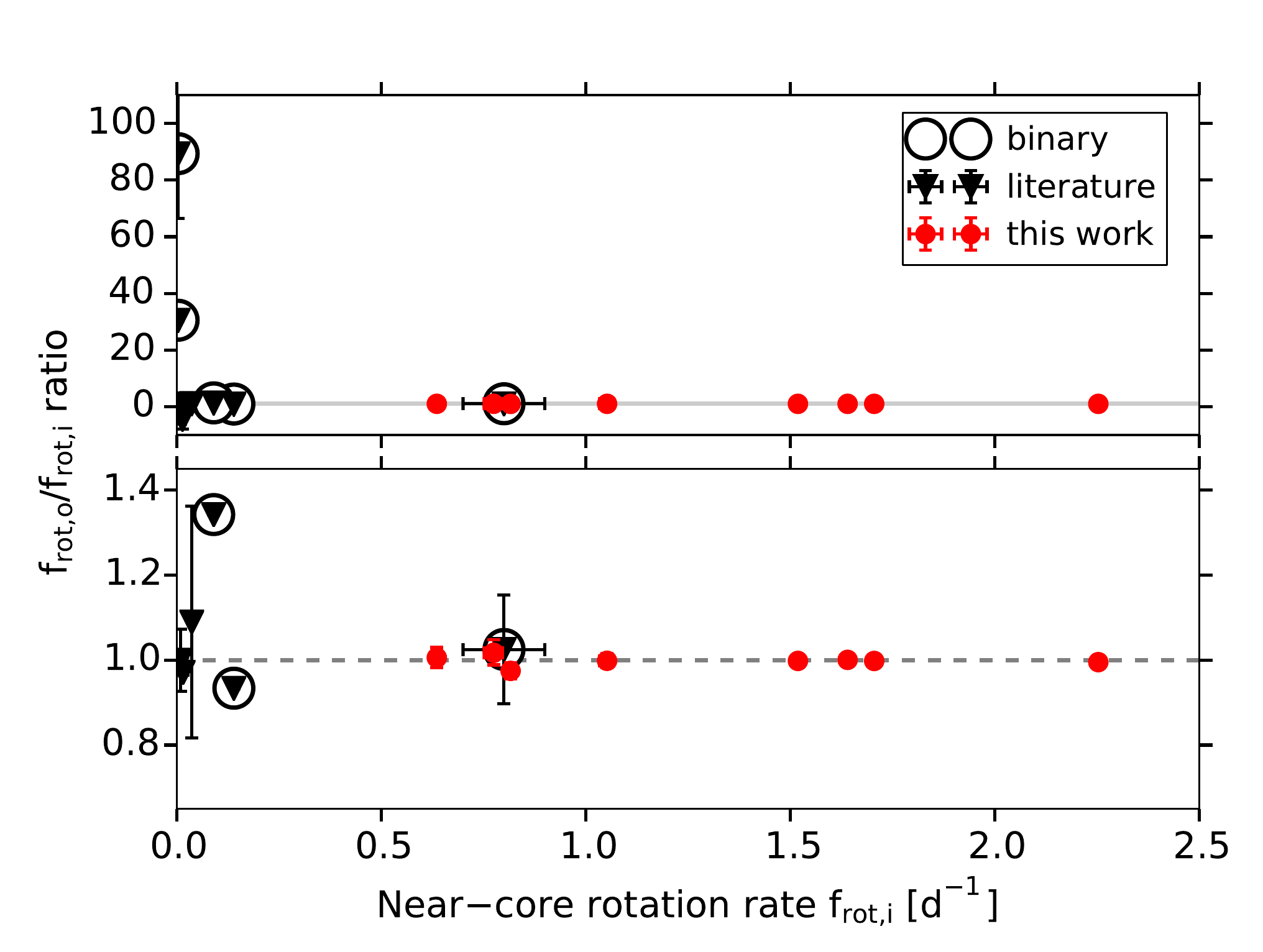}
 \caption{\label{fig:surfcore_overview} An overview of BAF-type main sequence stars with derived values for the near-core rotation rate $f_{\rm rot,i}$ and the surface rotation rate $f_{\rm rot,o}$. The bottom figure is a zoom-in of the top figure. For most stars, the values are consistent with rigid rotation within $1\sigma$. The observational signature of differential surface-to-core rotation becomes stronger when the star is slowly rotating, and/or in a binary system. The stars with values from the literature were taken from \citet{Kurtz2014,Saio2015,Triana2015,Moravveji2016,Murphy2016,SchmidAerts2016,Guo2017,Kallinger2017,Sowicka2017}.}
\end{figure}

Second, we have determined surface rotation rates for eight stars in our sample, using detected signal of rotational modulation. From these, we then derived the surface-to-core rotation ratio. Such values have already been computed for other stars in the literature, albeit for more slowly rotating stars. An overview of these stars is shown in Fig.\,\ref{fig:surfcore_overview}. As we can see, most stars are seemingly uniformly rotating within $1\sigma$ or $2\sigma$ error margins, and the differential rotation grows larger when {\em (i)} the near-core rotation rate is low, or {\em (ii)} the star is in a binary system, and sensitive to tidal effects.

When both methods for the detection and constraining of differential rotation in $\gamma$\,Dor stars are combined, we find that the observations are in agreement with (quasi-)rigid rotation throughout most of the radiative region, but that we cannot exclude differential rotation in the near-core regions. The precise location where such differential rotation may be present and the steepness of the gradient can strongly vary, and has to be considered for each studied star individually. When the stellar surface rotation rate is known as well, this can be included as a strong constraint in seismic modelling.

Finally, six stars in the sample exhibit a single retrograde period spacing pattern in qualitative agreement with Yanai modes, for $(k,m) = (-1,-1)$  \citep{Saio2018}. Considering the high rotation frequencies and the associated low inclination angles that this mode identification implies, two-dimensional models that do not assume the traditional approximation, are required to treat these pulsations properly.

This work provides us with a new observational window into the angular momentum transport that takes place in the interiors of intermediate- and high-mass stars with convective cores and radiative envelopes. We derived observational constraints for the rotation profiles in moderate to fast rotating stars, whereas available studies in the literature were mostly limited to slowly rotating stars. Consequently, the results of this work allow us to build a more complete picture of the stellar rotation. This is required to understand which angular momentum transport processes take place inside a star, resulting in the observed dramatic changes in stellar rotation rate during stellar evolution \citep[submitted]{Aerts2017, Ouazzani2018}.

%

\begin{acknowledgements}
  TVR is grateful for the kind hospitality and opportunity to perform part of this research at the Kavli Institute of Theoretical Physics, University of California at Santa Barbara, USA. The research leading to these results has received funding from the Research Foundation Flanders (FWO, Belgium, under grant agreements G.0B69.13 and V4.272.17N), from the European Research Council (ERC) under the
  European Union’s Horizon 2020 research and innovation programme (grant agreements No. 670519: MAMSIE, and No. 647383: SPIRE), from the National Science Foundation of the United States under grant NSF PHY-1748958, from the Belgian Science Policy Office (Belspo) under ESA/PRODEX grant ``PLATO mission development'', from the Fonds Wetenschappelijk Onderzoek - Vlaanderen (FWO) under the grant agreement G0H5416N (ERC Opvangproject), and from PLATO CNES grant at CEA-Saclay. We gratefully acknowledge (partial) support from the Australian Research Council, and from the Danish National Research Foundation (Grant DNRF106) through its funding for the Stellar Astrophysics Centre (SAC). AGH acknowledges funding support from Spanish public funds for research under project ESP2015-65712-C5-5-R (MINECO/FEDER), from project RYC-2012-09913 under the `Ram\'onn y Cajal' programme of the Spanish MINECO, from Funda\c{c}\~{a}o para a Ci\^encia e a Tecnologia (FCT, Portugal) through the fellowship SFRH/BPD/80619/2011 and from the European Council Project SPACEINN (FP7-SPACE-2012-312844). JG and RHDT acknowledge support from National Science Foundation grants AST 1716436 and ACI 1663696. We are grateful to Bill Paxton and his collaborators for their valuable work on the stellar evolution code MESA. Funding for the \emph{Kepler\hyphenation{Kep-ler}} mission is provided by NASA's Science Mission Directorate. We thank the whole team for the development and operations of the mission. This research made use of the SIMBAD database, operated at CDS, Strasbourg, France, and the SAO/NASA Astrophysics Data System. This research has made use of the VizieR catalogue access tool, CDS, Strasbourg, France.
\end{acknowledgements}

\bibliographystyle{aa} 
\bibliography{DiffRotation}

\begin{appendix}

\section{Gravito-inertial modes and seismic diagnosis in differentially rotating stars}
\label{Appendix:theor}

\subsection{The dynamics of low-frequency gravito-inertial modes within TAR}
To treat the dynamics of gravito-inertial modes in a differentially rotating star, we solve the inviscid system formed by the momentum equation
\begin{equation}
\left(\partial_{t}\!+\!\Omega\partial_{\varphi}\right)\vec u+2\Omega\,\widehat{\bf e}_{z}\times\vec u+r\sin\theta\left(\vec u\cdot\vec\nabla\Omega\right)\widehat{\bf e}_{\varphi}\!=\!
-\frac{1}{\overline\rho}\vec\nabla{\widetilde P}-\vec\nabla{\widetilde\Phi}+\frac{\widetilde\rho}{\overline\rho^2}\vec\nabla\overline P,
\end{equation}
the continuity equation $\left(\partial_{t}\!+\!\Omega\partial_{\varphi}\right){\widetilde\rho}+\vec\nabla\cdot\left(\overline\rho\vec u\right)=0$, the energy transport equation in the adiabatic limit
\begin{equation}
\left(\partial_{t}\!+\!\Omega\partial_{\varphi}\right)\left(\frac{\widetilde P}{\Gamma_1\overline P}-\frac{\widetilde\rho}{\overline\rho}\right)+\frac{N^2\left(r\right)}{\overline g}u_{r}=0
\end{equation}
and the Poisson's equation $\nabla^2\widetilde\Phi=4\pi G\widetilde\rho$ \citep{Unno1989}. $\rho$, $\Phi$, $P$ are respectively the fluid density, gravific potential and pressure. Each of them has been expanded as: $X\left(r,\theta,\varphi,t\right)={\overline X}\left(r\right)+{\widetilde X}\left(r,\theta,\varphi,t\right)$, where $\overline X$ is the mean hydrostatic value of $X$ on the isobar and $\widetilde X$ the wave's associated linear fluctuation. We introduce the angular velocity $\Omega\left(r,\theta\right)=2\pi f_{\rm rot}\left(r,\theta\right)$ and the Brunt-V\"ais\"al\"a frequency $N^2=\overline g\left(\frac{1}{\Gamma_1}\frac{{\rm d}\ln\overline P}{{\rm d}r}-\frac{{\rm d}\ln\overline\rho}{{\rm d}r}\right)$, where $\Gamma_1=\left(\partial\ln P/\partial\ln\rho\right)_{S}$ ($S$ being the macroscopic entropy) is the adiabatic exponent.  $\vec u$ is the wave velocity field. Finally, $(r,\theta,\varphi)$ are the usual spherical coordinates with their unit vector basis $\left\{\widehat{\bf e}_{j}\right\}_{j=\left\{r,\theta,\varphi\right\}}$ while $\widehat{\bf e}_{z}=\cos\theta\,\widehat{\bf e}_{r}-\sin\theta\,\widehat{\bf e}_{\theta}$ is the one along the rotation axis. $t$ is the time and $G$ the universal gravity constant.\\

To solve this system for low-frequency gravito-inertial modes, four main approximations can be assumed:

- {\it the Cowling approximation:} following \cite{Cowling1941}, the fluctuation of the gravitational potential (${\widetilde\Phi}$) can be neglected.

- {\it the anelastic approximation:} we filter out the high-frequency acoustic modes by simplifying the continuity equation that becomes $\vec\nabla\cdot\left({\overline\rho}{\vec u}\right)\approx 0$.

- {\it the JWKB approximation}: low-frequency modes are such that $\sigma\!<\!\!<\!N$ ($\sigma=2\pi f_{\rm in}$ is the wave frequency in an inertial reference frame). Therefore, we are studying modes which are rapidly oscillating along the radial direction and the JWKB approximation can be adopted. Each oscillating quantity is expanded as ${\widetilde X}\left(r,\theta,\varphi,t\right)\equiv{\widehat X}\left(r,\theta\right)\exp\left[i\int^{r}k_{V}\left(r'\right){\rm d}r'\right]\exp\left[i\left(\sigma t - m \varphi\right)\right]$ \citep{Ogilvie2004,Mathis2009}.

- {\it the Traditional Approximation for Rotation (TAR)}: stellar radiative zones are strongly stably stratified regions. In the case where the angular velocity ($\Omega$) is reasonably weak compared to the break-up angular velocity, $\Omega_{K}=\sqrt{G\,M/R^3}$, where $M$ and $R$ are the stellar mass and radius respectively, we can neglect the centrifugal acceleration at the first order. Moreover, if the stratification restoring force dominates the Coriolis acceleration along the radial direction (i.e. $2\Omega\!<\!\!<\!N$), this allows us to adopt the TAR where the latitudinal component (along $\widehat{\bf e}_{\theta}$) of the rotation vector $\vec \Omega=\Omega\,\widehat{\bf e}_{z}=\Omega\cos\theta\,\widehat{\bf e}_{r}-\Omega\sin\theta\,\widehat{\bf e}_{\theta}$ can be neglected for all latitudes while $u_{r}\!<\!\!<\!\left\{u_{\theta},u_{\varphi}\right\}$. We refer the reader to \cite{Friedlander1987}, \cite{Ogilvie2004} and \cite{Mathis2009} for detailed demonstrations.

Under those approximations, \cite{Mathis2009} derived the wave's velocity field:
\begin{equation}
\vec u\left(\vec r,t\right)=\sum_{j=\left\{r,\theta,\varphi\right\}}\left[\sum_{\sigma k m}u_{j;\sigma km}\left(\vec r,t\right)\right]\widehat{\bf e}_{j},
\label{uf}
\end{equation}
where
\vskip -24pt
\begin{equation}
u_{r;\sigma km}(\vec r,t)=\frac{\widehat\sigma}{N}\frac{{\widetilde\lambda}^{1/2}_{{\widehat\nu}km}\left(r\right)}{r}w_{\widehat\nu k m}\left(r,\theta\right)\!\sin\left[\Psi_{{\widehat\nu}km}\left(r,\varphi,t\right)\right]\!,
\label{urf}
\end{equation}
\vskip -20pt
\begin{eqnarray}
u_{\theta;\sigma km}(\vec r,t)&=&-\frac{\widehat\sigma}{r}{\mathcal G}_{\widehat\nu km}^{\theta}\left(r,\theta\right)\cos\left[\Psi_{{\widehat\nu}km}\left(r,\varphi,t\right)\right],
\label{utf}
\end{eqnarray}
\vskip -13pt
\begin{eqnarray}
u_{\varphi;\sigma km}(\vec r,t)&=&\frac{\widehat\sigma}{r}{\mathcal G}_{\widehat\nu km}^{\varphi}\left(r,\theta\right)\sin\left[\Psi_{{\widehat\nu}km}\left(r,\varphi,t\right)\right],
\label{upf}
\end{eqnarray}
with the Doppler-shifted local frequency ($\widehat\sigma$) and the spin parameter \citep[see][]{Lee1997}:
\begin{equation}
\widehat\sigma\left(r,\theta\right)=\sigma-m\Omega\left(r,\theta\right)\quad\hbox{and}\quad\widehat\nu\left(r,\theta\right)=\frac
{2\Omega\left(r,\theta\right)}{\widehat\sigma\left(r,\theta\right)}.
\end{equation}

Unlike the case of uniform rotation, variables do not separate neatly anymore within TAR in the case of general differential rotations $\Omega\left(r,\theta\right)$, even for shellular rotation laws $\Omega\left(r\right)$ \citep{Mathis2009}. The velocity components are thus expressed in terms of the 2D dynamical pressure ($P/\overline\rho$) eigenfunctions $w_{{\widehat\nu}km}$ which are solutions of the following eigenvalue equation:
\begin{equation}
{\mathcal O}_{\widehat\nu m}\left[w_{\widehat\nu k m}\left(r,x\right)\right]=-{\widetilde\lambda}_{\widehat\nu k m}\left(r\right)w_{\widehat\nu k m}\left(r,x\right),
\label{Eigen}
\end{equation}
where we have introduced the General Laplace Operator (GLO)
\begin{eqnarray}
{\mathcal O}_{\widehat\nu m}&=&\frac{1}{\widehat\sigma}\frac{{\rm d}}{{\rm d}x}\left[\frac{\left(1-x^2\right)}{\widehat\sigma{\mathcal D}_{\widehat\nu}\left(r,x\right)}\frac{{\rm d}}{{\rm d}x}\right]-\frac{m}{\widehat\sigma^2{\mathcal D}_{\widehat\nu}\left(r,x\right)}\left(1-x^2\right)\frac{\partial_{x}\Omega}{\widehat\sigma}\frac{{\rm d}}{{\rm d}x}
\nonumber\\
& &-\frac{1}{\widehat\sigma}\left[\frac{m^2}{\widehat\sigma{\mathcal D}_{\widehat\nu}\left(r,x\right)\left(1-x^2\right)}+m\frac{{\rm d}}{{\rm d}x}\left(\frac{\widehat\nu x}{\widehat\sigma{\mathcal D}_{\widehat\nu}\left(r,x\right)}\right)\right]
\end{eqnarray}
with
\begin{equation}
{\mathcal D}_{\widehat{\nu}}\left(r,x\right)=1-{\widehat\nu}^2x^2+{\widehat\nu}\,\left({\partial_{x}\Omega}/{\widehat\sigma}\right)x\left(1-x^2\right)
\label{EqD}
\end{equation}
and $x=\cos\theta$. The GLO is a differential operator in $x$ only with a parametric dependence on $r$ and the $w_{{\widehat\nu}km}$ form a complete orthogonal basis
\begin{equation}
\int_{-1}^{1}w^{*}_{{\widehat\nu}{k_{1}}m}\left(r,x\right)w_{{\widehat\nu}{k_2}m}\left(r,x\right){\rm d}x={\mathcal C}_{k_{1} m}\delta_{k_{1}k_{2}},
\end{equation}
where ${\mathcal C}_{k_{1} m}$ is the normalization factor and $\delta_{k_{1}k_{2}}$ is the usual Kronecker symbol. The use of the JWKB approximation allows us to reduce the problem of solving a 2D Partial Differential Equation to a Sturm-Liouville problem on co-latitude ($x$) for each radius ($r$). The GLO is the generalization of the classical Laplace tidal operator (Laplace 1799), its eigenfunctions $w_{{\widehat\nu}km}$ being thus a generalization of the Hough functions \citep{Hough1898,Ogilvie2004}\footnote{In the case of differential rotation, the possible latitudinal trapping of waves obtained for sub-inertial gravito-inertial waves in uniformly rotating stars will become more complex. We refer the reader to \citet{Mathis2009}, \citet{Mirouh2016} and \citet{Prat2018} for a detailed discussion.}. The ${\widetilde\lambda}_{\widehat\nu k m}\left(r\right)$ are its eigenvalues. The positive ones correspond to propagative modes for which the dispersion relation is given by
\begin{equation}
k_{V;\widehat\nu k m}^{2}\left(r\right)=\frac{{\widetilde\lambda}_{\widehat\nu k m} N^2}{r^2},
\label{dispRel}
\end{equation}
where $k_{V;\widehat\nu k m}$ is the vertical component of the wave vector. That leads to the following expressions for the JWKB phase function
\begin{equation}
\Psi_{\widehat\nu k m}\left(r,\varphi,t\right)=\sigma t+\int_{}^{r}k_{V;\widehat\nu k m}\,{\rm d}r^{'}+m\varphi.
\end{equation}
Finally, the latitudinal and azimuthal eigenfunctions are defined by
\begin{equation}
{\mathcal G}_{\widehat\nu km}^{\theta}\left(r,x\right)=\frac{1}{\widehat{\sigma}^{2}}\frac{1}{{\mathcal D}_{\widehat{\nu}}\left(r,x\right)\sqrt{1-x^2}}\left[-\left(1-x^2\right)\frac{{\rm d}}{{\rm d}x}+m\widehat\nu x\right]w_{{\widehat\nu}km}
\end{equation}
\vskip -13pt
\begin{eqnarray}
\lefteqn{{\mathcal G}_{\widehat\nu km}^{\varphi}\left(r,x\right)=\frac{1}{\widehat{\sigma}^{2}}\frac{1}{{\mathcal D}_{\widehat{\nu}}\left(r,x\right)\sqrt{1-x^2}}}\nonumber\\
&\times&\left[-\!\left(\widehat\nu x-\left(1-x^2\right)\frac{\partial_{x}\Omega}{\widehat\sigma}\right)\left(1-x^2\right)\!\frac{{\rm d}}{{\rm d}x}\!+\!m\right]w_{{\widehat\nu}km}.
\end{eqnarray}

\subsection{The case of a radial differential rotation: main properties and period spacing}
We now examine the case of a purely radial differential rotation
\begin{equation}
\Omega\left(r\right)=2\pi f_{\rm rot}\left(r\right).
\end{equation}
In this case, the local frequency and the local spin parameter become:
\begin{equation}
{\widehat\sigma}\left(r\right)=\sigma-m\Omega\left(r\right)\quad\hbox{and}\quad{\widehat\nu}\left(r\right)=\frac{2\Omega\left(r\right)}{\widehat\sigma\left(r\right)},
\end{equation}
where we recall that $\sigma=2\pi f_{in}$.

The GLO, ${\mathcal O}_{{\widehat \nu}m}$, then simplifies to the classical Laplace's tidal operator ${\mathcal L}_{{\widehat\nu}m}$:
\begin{equation}
   \begin{split}
      {\mathcal O}_{{\widehat \nu}m}&=\frac{1}{{\widehat\sigma}^{\,2}\left(r\right)}{\mathcal L}_{\widehat{\nu}m}\\
       &=\frac{1}{{\widehat\sigma}^{\,2}\left(r\right)}\Biggl[\frac{\rm d}{{\rm d}x}\left(\frac{1-x^2}{1-\widehat{\nu}^{\,2}\left(r\right) x^2}\frac{\rm d}{{\rm d}x}\right)\\
       &-\frac{1}{1-\widehat{\nu}^{\,2}\left(r\right) x^2}\left(\frac{m^2}{1-x^2}+m\widehat{\nu}^{\,2}\left(r\right)\frac{1+\widehat{\nu}^{\,2}\left(r\right)x^2}{1-\widehat{\nu}^{\,2}\left(r\right) x^2}\right)\Biggr]\,.\\
   \end{split}
\end{equation}
Its eigenfunctions are the usual Hough's functions $\Theta_{\widehat{\nu}km}\left(r,x\right)$ \citep{Hough1898} that depend both on $x$ and $r$ in the case of a radial differential rotation through the spin parameter ($\widehat\nu$) because it varies along the radius. In other words, we solve the classical Laplace tidal equation for each radius and corresponding values of the rotation angular velocity and spin parameter. In the particular case of a uniform rotation, $\widehat\nu$ is independent of $r$ and $x$. The Laplace's tidal operator thus becomes a linear differential operator in $x$ only with the corresponding purely latitudinal eigenfunctions \cite[e.g.][]{Lee1997,Townsend2003}.

The asymptotic expression of the radial component of the velocity field of low-frequency gravito-inertial modes becomes following Eq. (\ref{urf})
\begin{eqnarray}
u_{r;km}(\vec r,t)&=&{\mathcal A}\frac{1}{N}\frac{\lambda^{1/2}_{{\widehat\nu}km}\left(r\right)}{r}\frac{1}{k_{V;{\widehat\nu}km}^{1/2}\left(r\right)}\sin\left[\Psi_{{\widehat\nu}km}\left(r,\varphi,t\right)\right]\nonumber\\
&&\times\,\Theta_{{\widehat\nu}km}\left(r,x\right)
\end{eqnarray}
with
\begin{equation}
k_{V;{\widehat\nu}km}^{2}\left(r\right)=\frac{N^2}{{\widehat\sigma}^2}\frac{\lambda_{{\widehat\nu}km}\left(r\right)}{r^2}\quad\hbox{where}\quad\lambda_{{\widehat\nu}km}\left(r\right)={\widehat\sigma}{\widetilde\lambda}_{{\widehat\nu}km}\left(r\right).
\label{dispRelWDRSR}
\end{equation}
The eigenvalues for the Laplace's tidal operator $\left({\lambda}_{{\widehat\nu}km}\right)$ have been related to those of the GLO $\left({\widetilde\lambda}_{{\widehat\nu}km}\right)$. From now on, ${\widehat \nu}$ becomes $\nu$ in the equations to lighten the notations. We refer the reader to \cite{Mathis2009} for the details of the properties of the latitudinal and longitudinal components of the velocity field.
The last step is to write the radial quantification 
\begin{equation}2\pi^2\left(n+\alpha_g\right) = \int_{r_1}^{r_2}\frac{\sqrt{\lambda_{\nu km}\left(r\right)}}{f_{in}-mf_{\rm rot}\left(r\right)}\frac{N\left(r\right)}{r}\mathrm{d}r,\label{eq:Pg_diff_Appendix}\end{equation}
where $n$ is the radial quantum number, that provides the asymptotic frequencies and the period spacing.

\section{Gravito-inertial modes in the asymptotic regime: Taylor expansion}
\label{Appendix:coeff}
As discussed in Section \ref{subsec:theor}, we can describe the pulsation periods of the gravito-inertial modes in a differentially rotating star as
\begin{equation*}
 \Pi_0\left(n+\alpha_g\right) \approx \sqrt{\lambda_{s}}\left(P_{\rm co,s} + \sum_{i=1}^3a_iG_i\left(\nu\right)P_{\rm co,s}^{i+1}\right),
\end{equation*}
The coefficients $a_i$ in this expression are given by
\begin{eqnarray*}
 a_1 & = & f'_{{\rm rot},s}\langle r-r_s\rangle,\\
 a_2 & = & f''_{{\rm rot},s}\langle \left(r-r_s\right)^2\rangle,\\
 a_3 & = & \left(f'_{{\rm rot},s}\right)^2\langle \left(r-r_s\right)^2\rangle.
\end{eqnarray*}

The functions $G_i$ are given by
\begin{eqnarray*}
 G_1 & = & \frac{1}{\lambda_s}\left(\frac{\mathrm{d}\lambda}{\mathrm{d}\nu}\right)_s\left(1 + \frac{m\nu_s}{2}\right) + m,\\
 G_2 & = & \frac{1}{\lambda_s}\left(\frac{\mathrm{d}\lambda}{\mathrm{d}\nu}\right)_s\left(1 + \frac{m\nu_s}{2}\right) + 4m,\\
 G_3 & = & \left(\frac{2}{\lambda_s}\left(\frac{\mathrm{d^2}\lambda}{\mathrm{d}\nu^2}\right)_s - \frac{1}{\lambda^2_s}\left(\frac{\mathrm{d}\lambda}{\mathrm{d}\nu}\right)^2_s\right)\left(1 + \frac{m\nu_s}{2}\right)^2\\
 &&+ \frac{2m}{\lambda_s}\left(\frac{\mathrm{d}\lambda}{\mathrm{d}\nu}\right)_s\left(1 + \frac{m\nu_s}{2}\right) + 2m^2.
\end{eqnarray*}

\clearpage
\section{Simulated period spacing patterns}
 Here, we provide few additional stellar models for which we evaluate the impact of differential rotation on period spacing pattern to complete those already discussed in Sec. \ref{subsubsec:illustration}.
\begin{figure*}[h]
 \includegraphics[width=\textwidth]{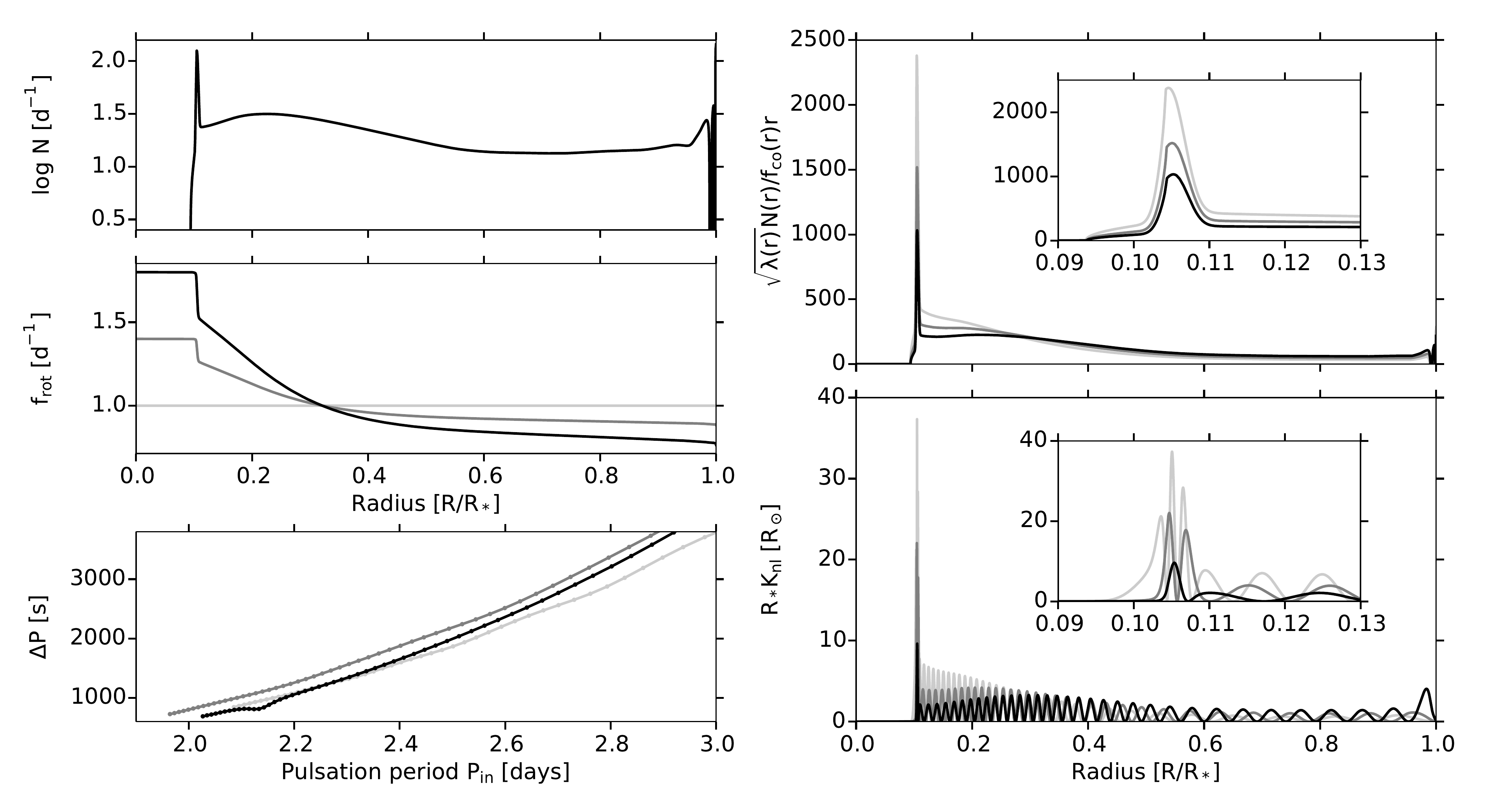}
\caption{\label{fig:yanaisim} The $1.7\,M_\odot$ stellar model at $X_c = 0.5$ with $f_{\rm rot,s} = 1\,\rm d^{-1}$ and $\delta f = 0.0$ (light grey), 0.4 (grey), and 0.8 (black), and the corresponding pulsations with $(k,m) = (-1,-1)$. {\em Top left:} the Brunt-V\"ais\"al\"a frequency profile. {\em Middle left:} The considered stellar rotation profiles. {\em Bottom left:} The period spacing patterns computed with GYRE v5.1 for the different rotation profiles. {\em Right}: The integrand of Eq.\,(\ref{eq:Pg_diff}) {(\em top)} and the rotation kernel {(\em bottom)} for the pulsation mode with $(n,k,m) = (-35,-1,-1)$.}
\end{figure*}

\begin{figure*}[h]
 \includegraphics[width=\textwidth]{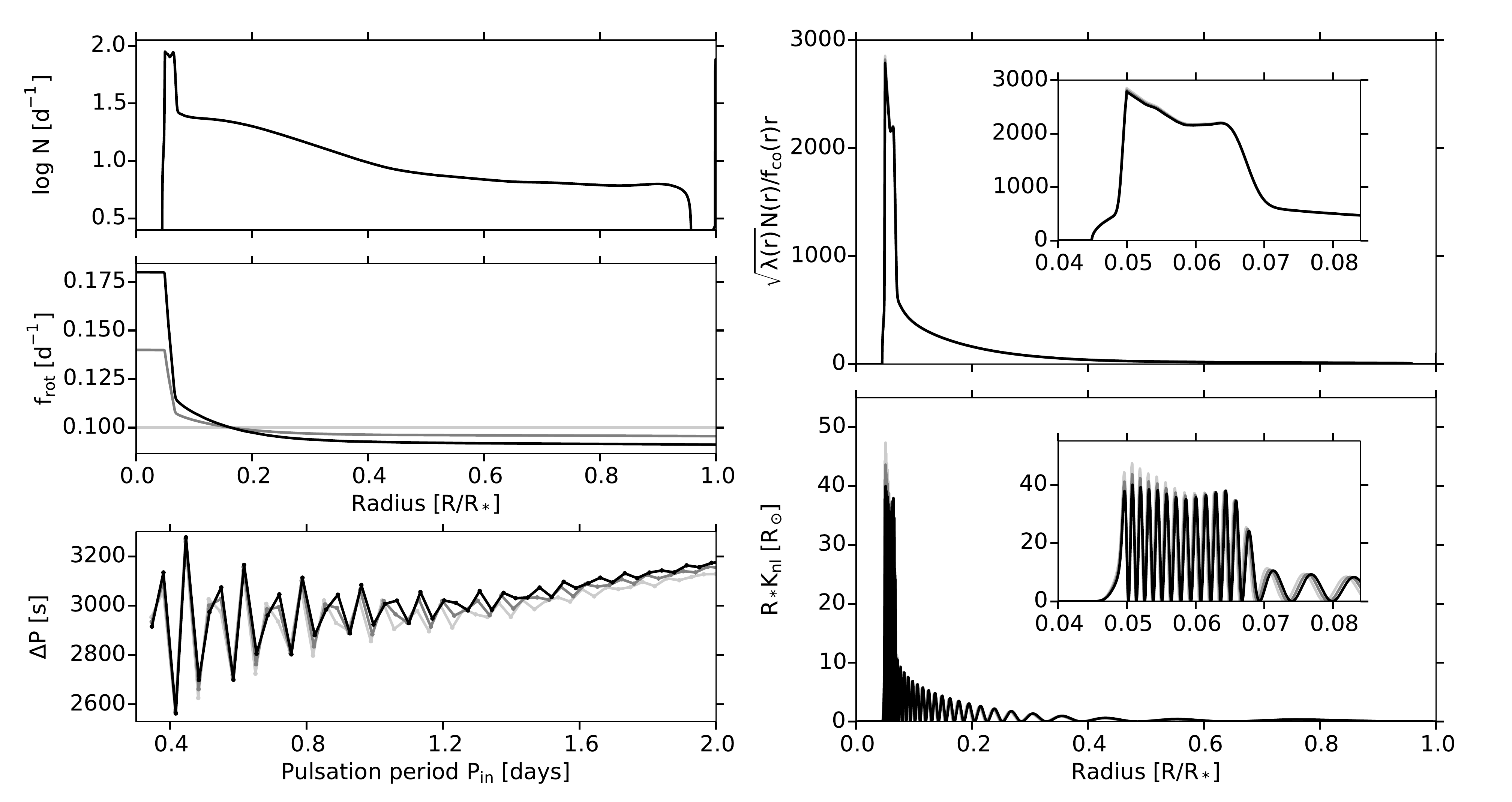}
\caption{\label{fig:retrogsim} The $1.7\,M_\odot$ stellar model at $X_c = 0.1$ with $f_{\rm rot,s} = 0.1\,\rm d^{-1}$ and $\delta f = 0.0$ (light grey), 0.4 (grey), and 0.8 (black), and the corresponding pulsations with $(k,m) = (+0,-1)$. {\em Top left:} the Brunt-V\"ais\"al\"a frequency profile. {\em Middle left:} The considered stellar rotation profiles. {\em Bottom left:} The period spacing patterns computed with GYRE v5.1 for the different rotation profiles. {\em Right}: The integrand of Eq.\,(\ref{eq:Pg_diff}) {(\em top)} and the rotation kernel {(\em bottom)} for the pulsation mode with $(n,k,m) = (-35,+0,-1)$.}
\end{figure*}
\end{appendix}

\end{document}